\documentclass[twocolumn]{aastex63}
\usepackage{amsmath}
\usepackage{rotating}
\usepackage{lineno}

\shorttitle{Seasonal Changes in HD~80606${\rm b}$'s Atmosphere}
\shortauthors{Sikora et al.}

\graphicspath{{./}{figures/}}

\begin{document}

\title{Seasonal Changes in the Atmosphere of HD~80606${\rm b}$ Observed with JWST's NIRSpec/G395H}

\correspondingauthor{James Sikora}
\email{james.t.sikora@gmail.com}

\author[0000-0002-3522-5846]{James T. Sikora}
\affiliation{Lowell Observatory, 1400 W Mars Hill Road, Flagstaff, AZ, 86001, USA}
\affiliation{Anton Pannekoek Institute for Astronomy, University of Amsterdam, 1098 XH Amsterdam, The Netherlands}
\affiliation{Department of Physics \& Astronomy, Bishop's University, Sherbrooke, QC J1M 1Z7, Canada}

\author[0000-0002-5904-1865]{Jason F. Rowe}
\affiliation{Department of Physics \& Astronomy, Bishop's University, Sherbrooke, QC J1M 1Z7, Canada}

\author[0000-0001-9987-467X]{Jared Splinter}
\affiliation{Trottier Space Institute at McGill, 3550 rue University, Montr\'eal, QC H3A 2A7, Canada}
\affiliation{Department of Earth and Planetary Sciences, McGill University, 3450 rue University, Montr\'eal, QC H3A 2T8, Canada}

\author[0009-0000-6113-0157]{Saugata Barat}
\affiliation{Anton Pannekoek Institute for Astronomy, University of Amsterdam, 1098 XH Amsterdam, The Netherlands}

\author[0000-0003-4987-6591]{Lisa Dang}
\affiliation{Institut Trottier de recherche sur les exoplanètes, Université de Montréal, 1375 Ave Thérèse-Lavoie-Roux, Montréal, QC, H2V 0B3, Canada}

\author[0000-0001-6129-5699]{Nicolas B. Cowan}
\affiliation{Trottier Space Institute at McGill, 3550 rue University, Montr\'eal, QC H3A 2A7, Canada}
\affiliation{Department of Physics, McGill University, 3600 rue University, Montr\'eal, QC H3A 2T8, Canada}
\affiliation{Department of Earth and Planetary Sciences, McGill University, 3450 rue University, Montr\'eal, QC H3A 2T8, Canada}

\author[0000-0001-7139-2724]{Thomas Barclay}
\affiliation{NASA Goddard Space Flight Center, 8800 Greenbelt Road, Greenbelt, MD 20771, USA}

\author[0000-0001-8020-7121]{Knicole D. Col\'on}
\affiliation{NASA Goddard Space Flight Center, 8800 Greenbelt Road, Greenbelt, MD 20771, USA}

\author[0000-0002-0875-8401]{Jean-Michel D\'esert}
\affiliation{Anton Pannekoek Institute for Astronomy, University of Amsterdam, 1098 XH Amsterdam, The Netherlands}

\author[0000-0002-7084-0529]{Stephen R. Kane}
\affiliation{Department of Earth and Planetary Sciences, University of California, Riverside, CA 92521, USA}

\author[0000-0000-0000-0000]{Joe Llama}
\affiliation{Lowell Observatory, 1400 W Mars Hill Road, Flagstaff, AZ, 86001, USA}

\author[0000-0000-0000-0000]{Hinna Shivkumar}
\affiliation{Anton Pannekoek Institute for Astronomy, University of Amsterdam, 1098 XH Amsterdam, The Netherlands}

\author[0000-0002-3481-9052]{Keivan G.\ Stassun}
\affiliation{Department of Physics and Astronomy, Vanderbilt University, Nashville, TN 37235, USA}

\author[0000-0003-1309-2904]{Elisa V. Quintana}
\affiliation{NASA Goddard Space Flight Center, 8800 Greenbelt Road, Greenbelt, MD 20771, USA}

\begin{abstract}
High-eccentricity gas giant planets serve as unique laboratories for studying the thermal and chemical properties of H/He-dominated atmospheres. One of the most extreme cases is HD 80606b---a hot Jupiter orbiting a sun-like star with an eccentricity of $0.93$---which experiences an increase in incident flux of nearly three orders of magnitude as the star-planet separation decreases from $0.88\,{\rm AU}$ at apoastron to $0.03\,{\rm AU}$ at periastron. We observed the planet's periastron passage using \emph{JWST}'s NIRSpec/G395H instrument ($2.8-5.2\,{\rm \mu m}$) during a $21\,{\rm hr}$ window centered on the eclipse. We find that, as the planet passes through periastron, its emission spectrum transitions from a featureless blackbody to one in which CO, CH$_4$, and H$_2$O absorption features are visible. We detect CH$_4$ during post-periapse phases at $4.1-10.7\sigma$ depending on the phase and on whether a flux offset is included to account for NRS1 detector systematics. Following periapse, H$_2$O and CO are also detected at $4.2-5.5\sigma$ and $3.7-4.4\sigma$, respectively. Furthermore, we rule out the presence of a strong temperature inversion near the IR photosphere based on the lack of obvious emission features throughout the observing window. General circulation models had predicted an inversion during periapse passage. Our study demonstrates the feasibility of studying hot Jupiter atmospheres using partial phase curves obtained with NIRSpec/G395H.
\end{abstract}


\section{Introduction}\label{sect:intro}

HD 80606b is a Jupiter-sized planet \citep[$R_{\rm p}=1.032\pm0.015\,R_{\rm Jup}$, $M_{\rm p}=4.1641\pm0.0047\,M_{\rm Jup}$;][]{pearson2022} orbiting a bright Sun-like star \citep[$J=7.702$, $T_{\rm eff}=5565\pm92\,{\rm K}$;][]{rosenthal2021}. It was initially discovered using radial velocity (RV) measurements obtained by \citet{naef2001} who reported a $111.8\,{\rm d}$ orbit with an eccentricity of $e=0.927$---the highest eccentricity of all previously detected exoplanets. \emph{Spitzer} IR photometry at $8.0\,{\rm \mu m}$ \citep{laughlin2009} was first used to detect the planet's eclipse occurring $\approx3\,{\rm hrs}$ prior to periapse. Subsequent photometric measurements revealed the planet's transit, occurring $\approx5.7\,{\rm d}$ after the eclipse \citep{fossey2009,garcia-melendo2009,moutou2009,winn2009,hebrard2010}. In-transit RV measurements have been used to detect the planet's Rossiter-McLaughlin effect, which suggests that its orbit may be misaligned with respect to the stellar rotation axis \citep{moutou2009,pont2009,hebrard2010}. Most recently, two of HD 80606b’s transits were detected using \emph{TESS} photometry \citep{ricker2014}, which, along with new RV measurements, provide the most precise constraints of the planet’s orbit to date \citep{pearson2022}.

\begin{figure}
	\centering
    \includegraphics[height=1\columnwidth]{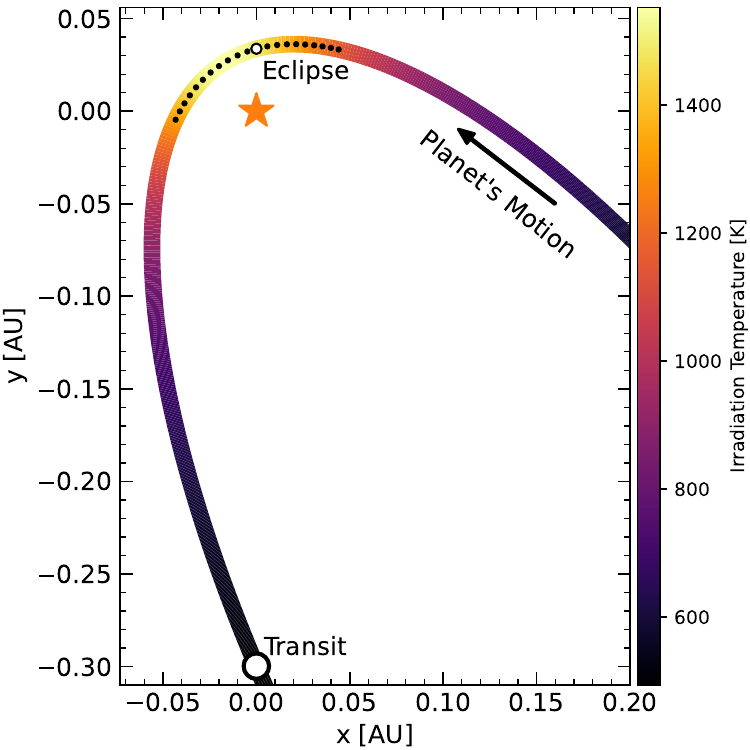}
	\caption{Black circles indicate the span of the NIRSpec observations plotted with $1\,{\rm hr}$ intervals. The color of the orbit indicates the irradiation temperature calculated based on the instantaneous star-planet separation and assuming zero albedo and full heat redistribution.}
	\label{fig:orbit}
\end{figure}

As a result of its high eccentricity, the stellar radiation that HD 80606b receives from its host star varies by a factor of $\approx860$ as it orbits from apoapse ($0.88\,{\rm AU}$) to periapse ($0.03\,{\rm AU}$) corresponding to an increase in irradiation temperature from $\approx300$ to $1500\,{\rm K}$ (calculated assuming full heat redistribution and zero albedo) (see Fig. \ref{fig:orbit}). This rapid change in incident flux, coupled with the planet’s favorable orbital configuration and bright host star, makes HD 80606b one of the best known laboratories for detecting and characterizing dynamical effects predicted to occur within the atmospheres of highly eccentric planets \citep[e.g.,][]{kane2009,iro2010,cowan2011,kane2011,mayorga2021,tsai2023}. 

Since HD 80606b was discovered, various models have been generated to study how its atmosphere behaves. Due to its relatively long orbital period, the vertical and horizontal temperature contrast throughout the planet’s atmosphere are expected to be low during most of the orbit \citep{langton2008,iro2010}. During the periapse passage, however, the dayside is rapidly heated, inducing strong acoustic waves \citep{langton2008} and accelerating winds \citep{lewis2017}. Depending on the radiative and advective timescales, the rapid heating may imprint a hotspot that periodically appears and disappears as the non-tidally locked planet rotates causing a ``ringing" to be observed in the thermal phase curve \citep{cowan2011,kataria2013}. It has been interpreted that the periastron passage also forms a transient dayside inversion characterized by a rise of $\sim400\,{\rm K}$ with decreasing pressure \citep{iro2010,mayorga2021,tsai2023}.

As the upper atmosphere gets heated while the planet orbits closer to the star, changes in the abundances and distributions of gasses and aerosols are also expected to occur. Assuming chemical equilibrium, methane (CH$_4$), which is predicted to dominate the atmosphere when the planet is less irradiated and cooler, may be converted into carbon monoxide (CO) near periastron \citep{visscher2012,tsai2023}. This process is complex and depends on the chemical timescale, efficiency of vertical mixing, strength, and timescale, and on the abundances of other species that are involved in the CO-CH$_4$ conversion. In addition, photochemical reactions \citep{moses2011,tsai2021} may deviate the atmospheric properties from chemical equilibrium. By coupling a 3D GCM with a 1D photochemical model, \citet{tsai2023} find that, as HD~80606b's CH$_4$ is converted into CO, the abundance of water (H$_2$O) also decreases while the abundances of certain photochemical products like hydrogen cyanide (HCN) and acetelyne (C$_2$H$_2$) increase well above what is expected under chemical equilibrium. Comparing condensation curves for various cloud species \citep{morley2012} with HD~80606b's temperature changes suggests that clouds may be present throughout much of the orbit. This was explored by \citet{lewis2017} using 3D GCMs, which demonstrate that clouds composed of MgSiO$_3$, MnS, and/or Na$_2$S likely persist even during periapse passage; these clouds may effectively raise the photosphere to lower pressures relative to that of a cloud-free atmosphere thereby changing the observed IR phase curve \citep{wit2016,lewis2017}.

All of the theoretical predictions developed for eccentric hot Jupiters in general and for HD 80606b in particular depend on certain assumptions about the planet’s rotation period, internal heating, and atmospheric composition. The first empirical insights into HD 80606b's atmospheric dynamics and thermal properties were derived using the $30\,{\rm hr}$ $8.0\,{\rm \mu m}$ and $80\,{\rm hr}$ $4.5\,{\rm \mu m}$ partial phase curves obtained with \emph{Spitzer} during the planet's periapse passage. \citet{laughlin2009} and \citet{wit2016} derived a radiative timescale of $\tau_{\rm rad}\sim4.5\,{\rm hrs}$ near the IR photosphere at atmospheric pressures $\sim10-100\,{\rm mbar}$. No ringing effect is detected in the $80\,{\rm hr}$ $4.5\,{\rm \mu m}$ planetary light curve \citep{wit2016}. This is consistent with the estimated rotation period of $P_{\rm rot}=93_{-35}^{+85}\,{\rm hr}$ derived by \citet{wit2016} using an energy-conserving semi-analytic atmospheric model (here, $P_{\rm rot}$ corresponds to the atmosphere's bulk rotation rate including winds and any underlying solid body rotation). HD 80606b's estimated $\tau_{\rm rad}$ and $P_{\rm rot}$ values are somewhat surprising considering theoretical predictions in which (1) cloud-free GCMs computed for HD 80606b have $\tau_{\rm rad}\sim8-12\,{\rm hrs}$ \citep{lewis2017} and (2) $P_{\rm rot}$ is expected to be $\sim40\,{\rm hr}$ under the assumption that the planet rotates at the pseudo-synchronous rotation period \citep{hut1981}.

While previous IR photometry of HD 80606b have provided important constraints on our understanding of the seasonal changes occurring in eccentric hot Jupiter atmospheres, wider wavelength coverage at higher spectroscopic resolutions is necessary in order to constrain predictions related to their chemical and thermal properties. The high mass of the planet makes HD 80606b a relatively poor target for atmospheric characterization via spectroscopy of the planet during transit. While there is potential evidence for an exosphere based on narrow-band transit spectroscopy \citep{colon2012}, eclipse observations in particular are necessary to learn more about the characteristics of HD 80606b's atmosphere.

Here we present near-IR spectroscopic measurements of HD 80606b’s periapse passage---including the eclipse---obtained with \emph{JWST}’s NIRSpec instrument. In Sect. \ref{sect:obs}, we describe the observations and the data reduction methods. In Sect. \ref{sect:analysis}, we present our analysis of both the integrated white light curves and the spectroscopic light curves. The results of our analysis are presented in Sect. \ref{sect:results} followed by a discussion and conclusions presented in Sections \ref{sect:disc} and \ref{sect:conclusions}.

\section{Observations}\label{sect:obs}

HD~80606 was observed using JWST's NIRSpec instrument in BOTS mode with the G395H/F290LP grating/filter combination as part of the GO-2488 program (PI:Sikora). A total of 13818 exposures were obtained over a $20.9\,{\rm hr}$ observing window from Nov. 1, 2022 at 06:25 UTC to Nov. 2, 2022 at 03:11 UTC. The observations were briefly interrupted during two $\approx5\,{\rm min}$ intervals for High Gain Antenna (HGA) moves: once during the eclipse (${\rm BJD}=2459885.164$) and once approximately $1\,{\rm hr}$ after periastron (${\rm BJD}=2459885.320$). The SUB2048 subarray was read-out using the NRSRAPID read-out pattern and we used five groups for each exposure (corresponding to a $5.43\,{\rm sec}$ exposure time) in order to avoid saturation for this bright target ($J=7.702\,{\rm mag}$).

Since the data presented in this work correspond to a partial rather than a full phase curve, we do not have a clear baseline that can be used to robustly remove certain systematic trends that are impacting the data. Therefore, carrying out independent reductions is necessary to assess how the systematics may be impacted by the reduction process. We carried out two entirely independent data reductions starting from the uncalibrated FITS files: one using the \texttt{Eureka!} Pipeline \citep{bell2022} and a second using a custom pipeline. We find that both reductions yield comparable results both in terms of the trends seen in the extracted planetary phase curve properties and in the systematic trends seen in the NRS1 and NRS2 measurements (presented in Sect. \ref{sect:SLCs}). We adopted the \texttt{Eureka!}-based reduction as the nominal one due to the lower scatter in the light curves and in the derived phase curve parameters associated with the spectral light curves.

\begin{figure*}
	\centering
	\includegraphics[width=0.98\textwidth]{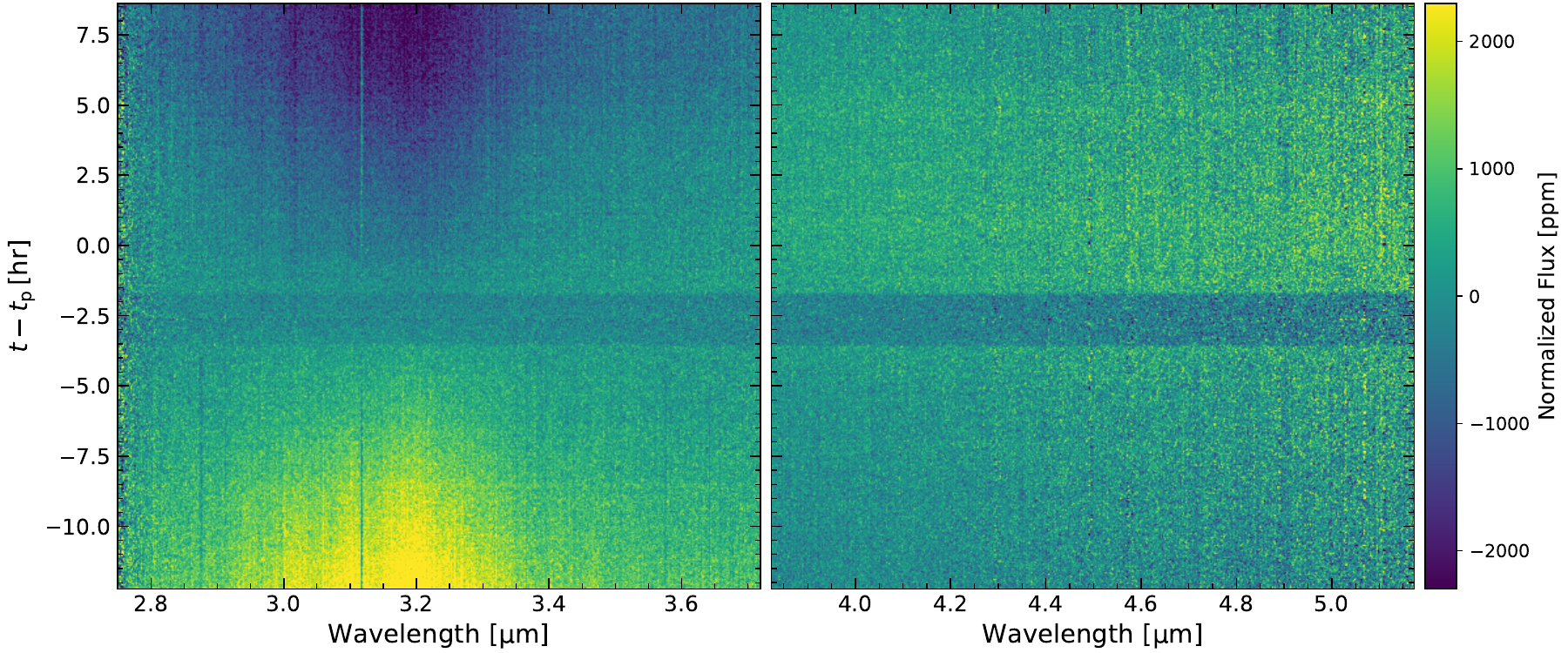}\vspace{-0.0cm}
    \caption{Observed NRS1 (left) and NRS2 (right) flux obtained using the \texttt{Eureka!} reduction pipeline and normalized by the median spectrum. Masked pixels were linearly interpolated over time and then over wavelength for easier visualization. The eclipse is visible near $t-t_{\rm p}=-2.6\,{\rm hr}$ in both detectors. The increase in the planet flux---peaking shortly after periapse---is also easily identifiable in the NRS2 detector. For the NRS1 spectra, systematic wavelength-dependent trends are apparent with the largest impact occurring near $3.2\,{\rm \mu m}$.}
    \label{fig:NRS1_NRS2_2D}
\end{figure*}

\subsection{Eureka! Reduction}\label{sect:eureka_reduction}

The \texttt{Eureka!} Pipeline \citep{bell2022} combines the first two stages of STScI's JWST Calibration Pipeline (\texttt{jwst}) \citep{bushouse2023} with a customized optimal spectral extraction routine. We used version 1.1 of the \texttt{Eureka!} Python package and 1.15.1 of the \texttt{jwst} Python package. Calibration reference files were taken from context 1303 of version 1.16.1 of the Calibration Reference Data System (CRDS). As briefly described below, we largely adopted the \texttt{Eureka!} settings for stages 1 to 3 that are recommended for NIRSpec/G395H BOTS mode observations while also making adjustments that were found to decrease the scatter in the white light curves. The settings that were used are defined in the \texttt{.ecf} control files, which we include in the online version of this paper for reference.

In Stage 1 of the pipeline, the measured ramps are fit to get the count rates for each exposure. Bad pixels identified using the `mask' reference FITS file from the CRDS were masked ($0.5\%$ of NRS1 pixels and $2.8\%$ of NRS2 pixels). Saturated pixels were flagged using the thresholds defined in the NRS1 and NRS2 `saturation' reference FITS files; these flags were then expanded using the \texttt{Eureka!} \texttt{update\_sat\_flags} keyword with \texttt{dq\_sat\_percentile} set to 50 and the \texttt{expand\_prev\_group} option enabled. An average of 9 pixels per integration ($0.01\%$) were found to be saturated for NRS1 and 162 pixels per integration ($0.2\%$) were saturated for NRS2. Group level column-by-column background subtraction was carried out by fitting a zeroth order polynomial to the counts in those pixels that are at least 8 pixels away from the trace (\texttt{masktrace} set to \texttt{True} and \texttt{expand\_mask} set to 8) where a $6\sigma$ threshold was adopted to remove outliers. We skipped the photometric calibration step in Stage 2 because it allows for a straight forward calculation of the photon noise limit using the electron counts.

In Stage 3, \texttt{Eureka!} performs an optimal spectral extraction routine. For NRS1, the spectra were extracted across columns 610 through 2044 while for NRS2, we used columns 5 through 2040. Outliers rejection along the time axis was carried out for each pixel using a $6\sigma$ rejection threshold. A Gaussian fit was used to identify the position of the source along each column (\texttt{src\_pos\_type=gaussian}). We found that super-sampling each centroid by a factor of 3 using the \texttt{super\_sample} parameter helped to reduce the scatter in the measured centroid positions and in the extracted white light curve. Background subtraction was carried out by removing the median flux calculated from pixels $\geq8$ pixels away from the source position while removing $>6\sigma$ outliers (\texttt{p3thresh=6}). The spectral extraction used an aperture with a half-width of 4 pixels.  We also tried using a half-width of 2 pixels but this led to a significantly higher white light curve scatter. Outliers rejection was carried out as part of steps 5 and 7 of the optimal spectral extraction routine using \texttt{p5thresh=6} and \texttt{p7thresh=6}. Thirteen integrations potentially affected by the 2 HGA moves noted in Sect. \ref{sect:obs} were manually removed across all wavelength channels along with one NRS1 light curve that was removed due to anomalous flux values, which can be seen in Fig. \ref{fig:NRS1_NRS2_2D} as the column near $3.1\,{\rm \mu m}$ showing lower/higher pre-/post-eclipse flux relative to adjacent columns. Sigma clipping was then carried out for each detector's $1434/2035$ wavelength channels with $6\sigma$ outliers being removed (1239 measurements in total for NRS1 and 2080 measurements for NRS2 or $<0.01\%$).

White light curves and spectral light curves were generated by binning the flux values extracted from Stage 3 across each wavelength channel bin, where the spectral wavelength bins are defined using a $0.1\,{\rm \mu m}$ bin width ($R\sim40$) resulting in 10 bins for NRS1 and 14 bins for NRS2. Finally, we carried out an additional iteration of sigma clipping for each of the wavelength-binned light curves. This did not remove any flux measurements from the NRS1 white/spectral light curves while 2 points for the NRS2 white light curve were removed and 8-17 points were removed from each of the NRS2 spectral light curves.

The NRS1 and NRS2 white light curves have root-mean-square (RMS) scatter of $163\,{\rm ppm}$ and $217\,{\rm ppm}$, respectively. In Fig. \ref{fig:NRS1_NRS2_2D}, we show the flux obtained from the \emph{Eureka!} pipeline normalized by the median spectrum. The eclipse (occurring $\approx2.6\,{\rm hr}$ before periapse) is visible in both the NRS1 and NRS2 detectors. The rise in the planet's flux following the eclipse is also clearly visible in the NRS2 detector but is somewhat obscured in NRS1 due to significant wavelength-dependent systematic trends, particularly at wavelengths $<3.5\,{\rm \mu m}$.

\begin{table}
       \caption{Published stellar and planetary properties associated with HD~80606b that were used in this study.}\vspace{-0.5cm}
       \label{tbl:pub_param}
       \begin{center}
       \begin{tabular}{@{\extracolsep{\fill}}l @{\hskip 1.cm} r @{\extracolsep{\fill}}}
              \hline
              \hline
              \noalign{\vskip0.5mm}
Parameter & Value \\
\hline
$T_{\rm eff}\,[{\rm K}]$           & $5565\pm92^{\rm a}$ \\
$R_\star\,[R_\odot]$               & $1.066\pm0.024^{\rm a}$ \\
$M_\star\,[M_\odot]$               & $1.047\pm0.047^{\rm a}$ \\
$\log g_\star\,[{\rm cm/s^2}]$     & $4.402\pm0.039^{\dagger}$ \\
$[{\rm Fe/H}]_\star$               & $0.348\pm0.057^{\rm a}$ \\
\hline
$M_{\rm p}\,[M_{\rm Jup}]$ & $4.1641\pm0.0047^{\rm b}$ \\
$R_{\rm p}\,[R_{\rm Jup}]$ & $1.032\pm0.015^{\rm b}$ \\
$(R_{\rm p}/R_\star)^2$    & $0.01019\pm0.00023^{\rm b}$ \\
$P_{\rm orb}\,[{\rm d}]$   & $111.436765\pm0.000074^{\rm b}$ \\
$a\,[{\rm au}]$            & $0.4603\pm0.0021^{\rm b}$ \\
$e$                        & $0.93183\pm0.00014^{\rm b}$ \\
$\omega\,[{\rm deg}]$      & $-58.887\pm0.043^{\rm b}$ \\
$T_{\rm tr}\,[{\rm BJD}-2400000]$ & $58888.07466\pm0.00204^{\rm b}$ \\
\noalign{\vskip0.5mm}
\hline
\multicolumn{2}{l}{$^{\rm a}$\citet{rosenthal2021}, $^{\rm b}$\citet{pearson2022}} \\
\multicolumn{2}{l}{$^\dagger$Calculated from the published $M_\star$ and $R_\star$.} \\
       \end{tabular}
       \end{center}
\end{table}

\subsection{Custom Pipeline}\label{sect:custom_reduction}

A fully custom, independent pipeline was written to validate the analysis from Sect. \ref{sect:eureka_reduction}.  The pipeline corrects for instrumental effects and extracts the spectro-photometry using difference imaging techniques. Non-linearity, superbias corrections are made using identical reference pixels as described in Sect. \ref{sect:eureka_reduction}.  Reference pixels corrections used simple means with 3$\sigma$ rejection for outliers.  The pipeline starts with Stage 0 data products (e.g., ``uncal.fits") and uses the saturation map reference files for NRS1 and NRS2 to flag any group during an exposure that exceeds saturation.  Saturated group values for each pixel  were excluded when integrating ``up-the-ramp" to estimate the electron counts for each pixel for each integration.  Superbias, reference pixel and non-linearity correction are then applied.  Bad pixels were flagged through saturation or non-zero values of bit 0 from the data-quality flags.  Bad-pixels were replaced by the mean of a 3x3 box centred on the offending group value.  

The pipeline uses a two-step process to calculate count rates and to correct for ``1/f" noise using difference images.  The first step uses linear-regression to estimate residual bias ($\it zpt$) and the integrated electron count rate.  Column and row means of the $\it zpt$ values for each integration were computed and then used as an additional bias correction for each group image.  A stacked median image for each group step was calculated using all valid integrations.  This provides five 32x2048 deep stacks based on 13,818 integrations.  Difference images for each group set are then used to correct for 1/f noise using column medians.  Only pixels more that 9 pixels away from the spectral trace were used when calculating the median.  The 1/f corrected groups were used to then calculate count rates based on a robust mean of forward-differences, with outliers excluded at the 2-$\sigma$ level. 

A median image was then used to calculate difference time-series images.  Aperture photometry along the spectral trace using an aperture radius of 4 pixels was used to extract the difference spectrum using both difference and non-difference images.  The median of the spectro-photometry from non-difference images was used to estimate the photometric zero-point to calculate the scaling necessary to convert differential counts to relative spectro-photometry.  

We estimate $200\,{\rm ppm}$ and $268\,{\rm ppm}$ RMS scatter for the NRS1 and NRS2 white light curves obtained from this custom reduction pipeline, respectively; these correspond to $1.2\times$ the RMS scatter obtained from the \emph{Eureka!} reduction.

\begin{figure}
	\centering
    \includegraphics[width=1\columnwidth]{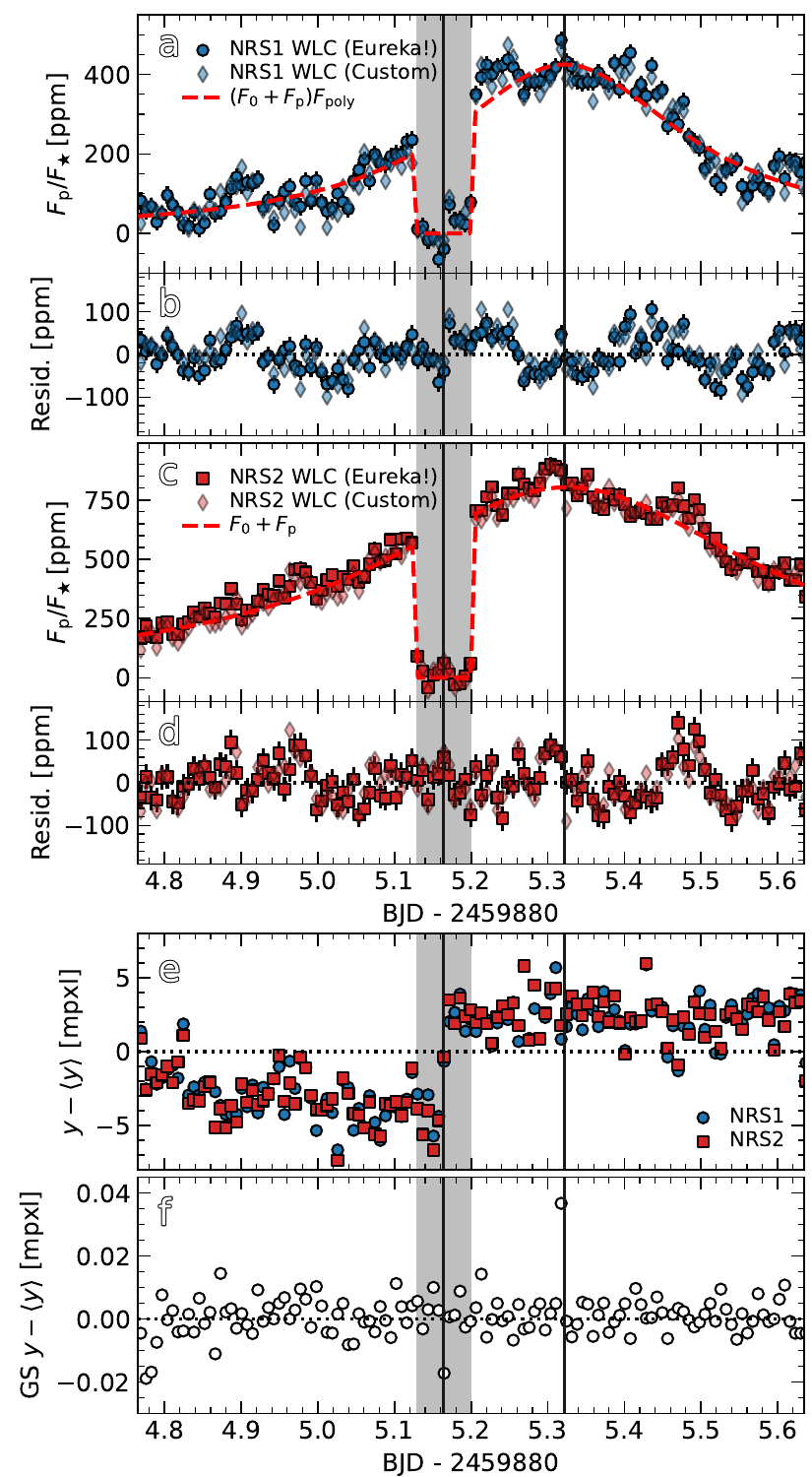}
	\caption{Binned NRS1 and NRS2 white light curves and their residuals (panels `a' to `d') obtained using a model that includes only the planet phase curve and, in the case of NRS1, a second order polynomial term (red dashed line). The darker circles/squares are from the Eureka! reduction and the fainter diamonds are from the independent custom reduction, which have comparable error bars but are not shown for visual clarity. Panels `e' and `f' show the relative centroid positions obtained from the Eureka! NRS1 and NRS2 reductions and the centroid positions associated with the guide star. All points have been binned using $10\,{\rm min}$ bin widths. The shaded region highlights the in-eclipse measurements and the vertical black lines indicate the two HGA moves.}
	\label{fig:centroid}
\end{figure}

\section{Analysis}\label{sect:analysis}

\subsection{Analytic Light Curve Model}\label{sect:LC_model}

We fitted the light curves obtained from the NRS1 and NRS2 detectors using an analytic model consisting of five components: (1) a planet flux ($F_{\rm p}$), (2) a constant stellar flux ($F_0$), (3) an additive sinusoidal term to remove flux variations caused either by the star or by instrumental systematics ($F_{\rm sin}$), (4) a multiplicative polynomial term ($F_{\rm poly}$), and (5) a multiplicative systematics term ($F_{\rm sys}$) such that
\begin{equation}\label{eqn:Ftot}
    F_{\rm obs}(t)=\Biggl(F_0+F_{\rm sin}(t)+F_{\rm p}(t)\biggr)F_{\rm sys}(t)F_{\rm poly}(t).
\end{equation}
For the planet flux, we adopted the asymmetric Lorentzian model proposed by \citet{lewis2013} in their analysis of Spitzer phase curve measurements of the eccentric hot Jupiter HAT-P-2b. The model has four parameters corresponding to the phase curve amplitude ($c_1$), which we define in terms of $F_0$, the peak offset time ($c_2$), and the timescales over which the flux rises ($c_3$) and decays ($c_4$) with respect to the phase curve peak:
\begin{equation}\label{eqn:Fp}
    F_{\rm p}(t)=F_0\frac{c_1}{u(t)^2+1}
\end{equation}
\begin{equation}
  {\rm where}\;\;\;u(t)=\begin{cases}
    (t-c_2)/c_3, & \text{if $t<c_2$}\\
    (t-c_2)/c_4, & \text{if $t>c_2$}.
  \end{cases}
\end{equation}
The symmetric version of this model corresponds to the case in which $c_3=c_4$. The eclipse was modeled using the \texttt{batman} Python package \citep{kreidberg2015} and depends on the orbital period ($P_{\rm orb}$), eccentricity ($e$), inclination angle ($i$), argument of periastron ($\omega$), eclipse mid-point ($T_{\rm e}$), the ratio of the semi-major axis to the stellar radius ($a/R_\star$), and the ratio of the planetary radius to $R_\star$ ($R_{\rm p}/R_\star$). Since the partial phase curve observations we obtained span only a small fraction of the $111.4\,{\rm d}$ orbit and do not include the transit, we opted to fix $P_{\rm orb}$, $e$, $i$, $\omega$, $a/R_\star$, and $R_{\rm p}/R_\star$ at published values (see Table \ref{tbl:pub_param}), which have been precisely measured using transit, eclipse, and radial velocity measurements \citep{pearson2022}. The eclipse mid-point, however, was allowed to vary as a free parameter when fitting the integrated white light curves.

The NRS1 measurements are dominated by a systematic downward slope in flux between the start and end of the observing window that is strongly wavelength-dependent (Fig. \ref{fig:NRS1_NRS2_2D}). Similar slopes have previously been reported for NIRSpec time series observations \citep[e.g.,][]{alderson2023,moran2023}. We account for these variations by including a multiplicative polynomial term,
\begin{equation}\label{eqn:Fpoly}
    F_{\rm poly}(t)=1+\sum_n p_n(t-t_0)^n
\end{equation}
where $p_n$ are free parameters and $t_0$ is the approximate eclipse mid-point time, which we fix at ${\rm BJD}=2459885.165$. No such systematic slope is visible by eye in the NRS2 white light curve, however we include the $F_{\rm poly}$ term in the model selection process described below.

In Fig. \ref{fig:centroid}, the panels labeled `a' to 'd' show the NRS1 and NRS2 white light curve flux and residuals obtained using the simplest model that produces an adequate overall fit to the data (discussed further in Sect. \ref{sect:WLCs}). For NRS1, this model consists of the star and planet contributions ($F_0$ and a symmetric $F_{\rm p}$) and a second order $F_{\rm poly}$ term while NRS2 only includes $F_0$ and $F_{\rm p}$. The residuals show correlated noise that we attempt to capture with the $F_{\rm sys}$ and $F_{\rm sin}$ terms. The sinusoidal term is defined as
\begin{equation}\label{eqn:fstar}
    F_{\rm sin}(t)=F_0(A_1+B_1[t-t_{\rm ini}])\sin{(2\pi[tf_1+\phi_1])}
\end{equation}
where $A_1$ and $B_1$ are free parameters included so as to allow the total amplitude to vary linearly over time and $t_{\rm ini}$ is the timestamp of the first exposure. The phase $\phi_1$ is included as a free parameter and the frequency $f_1$ is fixed at the highest-amplitude frequencies identified in Lomb Scargle periodograms \citep{lomb1976,scargle1982} calculated from the residuals shown in Fig. \ref{fig:centroid}. These periodograms exhibit maximum amplitude peaks at frequencies of $0.174\,{\rm hr^{-1}}$ ($5.7\,{\rm hr}$) and $0.231\,{\rm hr^{-1}}$ ($4.3\,{\rm hr}$) for the NRS1 and NRS2 light curves, respectively.

The bottom two panels of Fig. \ref{fig:centroid} (`e' and `f') show the NRS1 and NRS2 centroid positions in the dispersion direction ($y$) obtained from the \texttt{Eureka!} reduction as well as the guide star centroid $y$ position ($y_{\rm GS}$). We find that both $y(t)$ and $y_{\rm GS}(t)$ show variability most notably during the first HGA move that occurs near the middle of the eclipse. The NRS1 residuals show a jump in flux for both the \texttt{Eureka!} reduction and for the custom reduction that also coincides with the first HGA move while no similar jump is apparent in the NRS2 residuals. This suggests that the measured NRS1 flux may be correlated with the centroid position due to intrapixel variations as found in \emph{Spitzer} and \emph{HST} time series observations \citep[e.g.,][]{desert2009,sing2019}. To account for this jump in flux and any other correlations with $y$ and $y_{\rm GS}$, we include an $F_{\rm sys}$ term defined as
\begin{align}\label{eqn:Fsys}
    F_{\rm sys}=1+c_yy^\prime(t)+c_{{\rm GS},y}y_{\rm GS}(t)+\Delta F_{\rm jump}(t),
\end{align}
where $c_y$ and $c_{{\rm GS},y}$ are free parameters. $y^\prime(t)$ corresponds to $y(t)$ after removing the jump, which is done by splitting the time series at the first HGA move and dividing the two segments by their median values. Both $y^\prime(t)$ and $y_{\rm GS}(t)$ are normalized by subtracting the mean and dividing by the standard deviation. The jump in flux is corrected using a step function $\Delta F_{\rm jump}(t)$ given by
\begin{equation}
  \Delta F_{\rm jump}(t)=\begin{cases}
    0, & \text{if $t<T_{\rm jump}$}\\
    \Delta F_{\rm jump}, & \text{if $t\geq T_{\rm jump}$}
  \end{cases}
\end{equation}
where $\Delta F_{\rm jump}$ and $T_{\rm jump}$ are free parameters. In Sect. \ref{sect:results}, we discuss how the nominal models for the NRS1 and NRS2 light curves were ultimately selected.

The best fitting parameters and uncertainties for the analytic light curve modelling were derived using the \texttt{emcee} ensemble sampler Python package \citep{foreman-mackey2013,foreman-mackey2019}. For each light curve, 24 to 48 walkers were initialized, generating chains $\gtrsim5\times10^4$ steps in length after discarding the first $5000$ steps as burn-in. Sampling was continued until the number of steps exceeded 50 times the mean auto-correlation lengths. The $\hat{R}$ statistic \citep{gelman1992} was then used to check for convergence where additional samples were generated until $\hat{R}<1.01$ for each parameter. For both the white light curve fits and the spectral light curve fits, we used uniform priors for all parameters. In all cases, we also included a `jitter' term ($\sigma_{\rm Jit}$, sampled in log space) that is added to the formal measurement uncertainties in quadrature, which accounts for additional white noise that may be present in the data. This increase in the measurement errors ultimately imposes a reduced $\chi^2$ of $\approx1$.

\subsection{Atmospheric retrievals}\label{sect:retrievals}

We carried out two sets of atmospheric retrievals using two independent frameworks. Within one framework, we used forward models calculated with \texttt{petitRADTRANS} \citep{molliere2019} in which the atmosphere is assumed to be in chemical equilibrium. The chemical composition of the atmosphere is determined by the bulk metallicity ([M/H]) and C/O ratio. The abundances are interpolated from a grid computed using \texttt{easyCHEM} \citep[for more details, see][]{molliere2017} and do not incorporate photochemical reactions. We included opacities from gas-phase H$_2$O, CO, CO$_2$, and CH$_4$, which are expected to be the predominant contributors to the total atomic and molecular opacity within the NRS1/NRS2 bandpasses. Continuum opacities associated with H$_2$-H$_2$ and H$_2$-He collision-induced absorption and H$_2$ and He Rayleigh scattering were also included. Clouds composed of MnS and/or MgSiO$_3$ could potentially be present during the observed phases \citep{lewis2017}. In order to account for these species, we include a gray cloud deck in which the flux emitted by layers below the cloud deck pressure ($P_{\rm cloud}$) is blocked at all wavelengths. The model atmosphere was defined using 100 layers logarithmically distributed from $100\,{\rm bar}$ to $1\,{\rm \mu bar}$ with a temperature described using the 4 parameter profile proposed by \citet{guillot2010} described by the IR absorption coefficient ($\kappa_{\rm IR}$), the ratio of optical to IR absorption coefficients ($\gamma$), the planet's interior temperature ($T_{\rm int}$), and the irradiation temperature ($T_{\rm irr}$).Posterior distributions were derived using \texttt{emcee} in which 16 walkers were initialized and $10^5$ samples per walker were generated with the first $10^4$ steps being discarded as burn-in. After removing chains stuck in low-probability regions (between 0 and 5 chains), we obtained $\sim10^6$ samples. Uniform priors were adopted for all parameters.

The second retrieval framework used forward models calculated with \texttt{PyratBay} \citep{cubillos2021} in which we include all of the opacity sources used in the \texttt{petitRADTRANS}-based retrievals listed above. In this case, the abundances are uniform with pressure and are allowed to vary freely. We used the same PT profile, cloud deck and adopted a model atmosphere of 61 layers with a minimum pressure at the top of the atmosphere of $100\,\mu{\rm bar}$. We sample parameter space with \texttt{MC3} using the Snooker Differential-Evolution MCMC algorithm \citep{Cubillos2017-mc3}. We initialize with 32 parallel chains, a total of $4 \times10^6$ samples with $10^5$ discareded as burn-in. We similarly adopt uniform priors for all parameters.

\begin{figure}
	\centering
	\includegraphics[width=1\columnwidth]{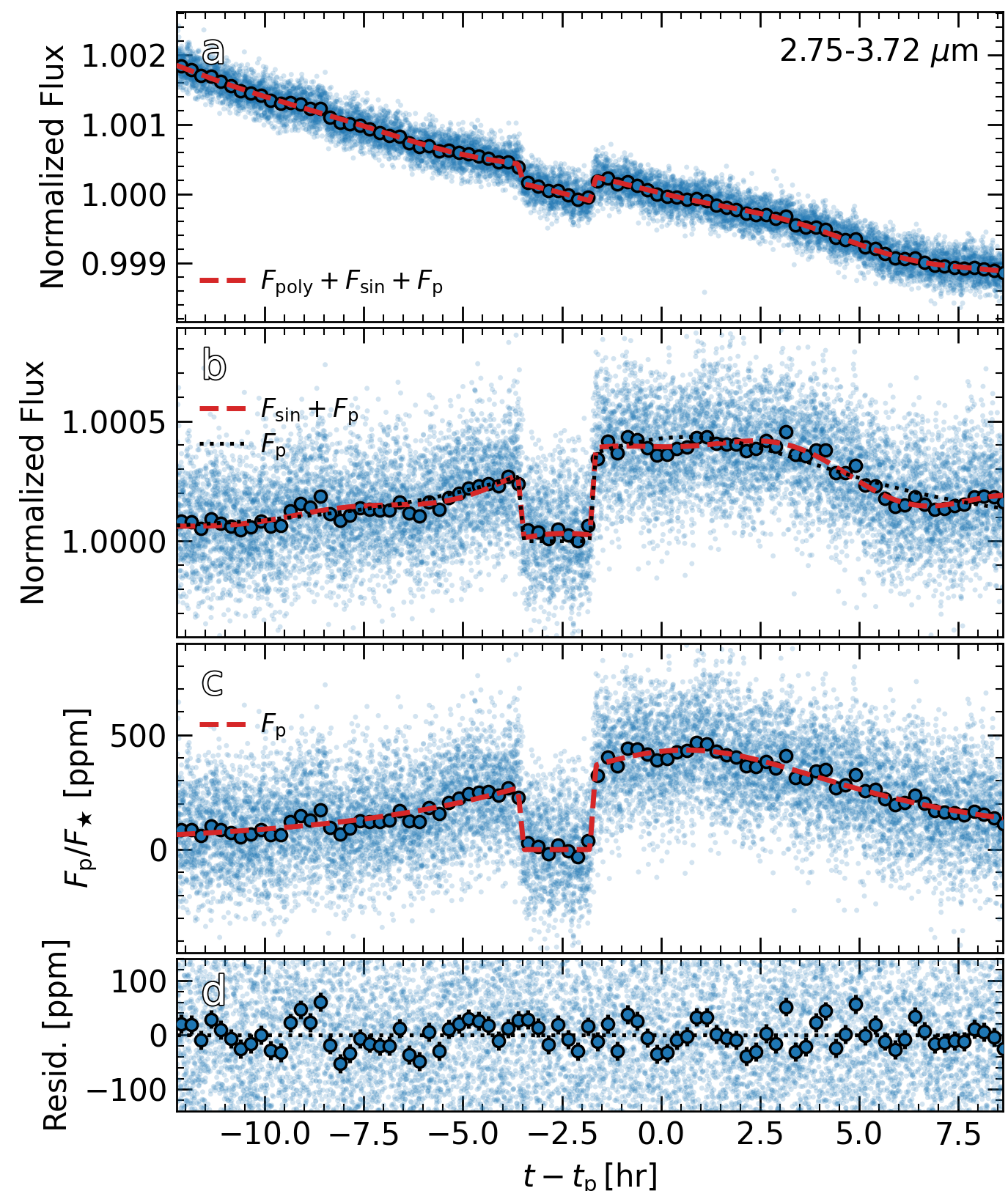}\vspace{-0.0cm}
    \caption{NRS1 white light curves where the dark blue circles are the binned measurements generated using $15\,{\rm min}$ bin widths and the light blue circles are the unbinned measurements. The $x$-axis corresponds to time relative to periapse. Panel `a' shows the observed flux obtained using \texttt{Eureka!}. The red dashed line is the best-fitting nominal model (Sect. \ref{sect:WLCs}) consisting of a symmetric planet phase curve, a sinusoidal term ($F_{\rm sin}$) with a frequency of $f_1=0.174\,{\rm hr^{-1}}$, and a second order polynomial term ($F_{\rm poly}$). In panel `b' the observations have been detrended by removing $F_{\rm poly}$, in panel `c' both $F_{\rm poly}$ and $F_{\rm sin}$ have been removed leaving the planetary signal ($F_{\rm p}$), and the residuals are shown in panel `d'.}
    \label{fig:NRS1_WL}
\end{figure}

\begin{figure}
	\centering
    \includegraphics[width=1\columnwidth]{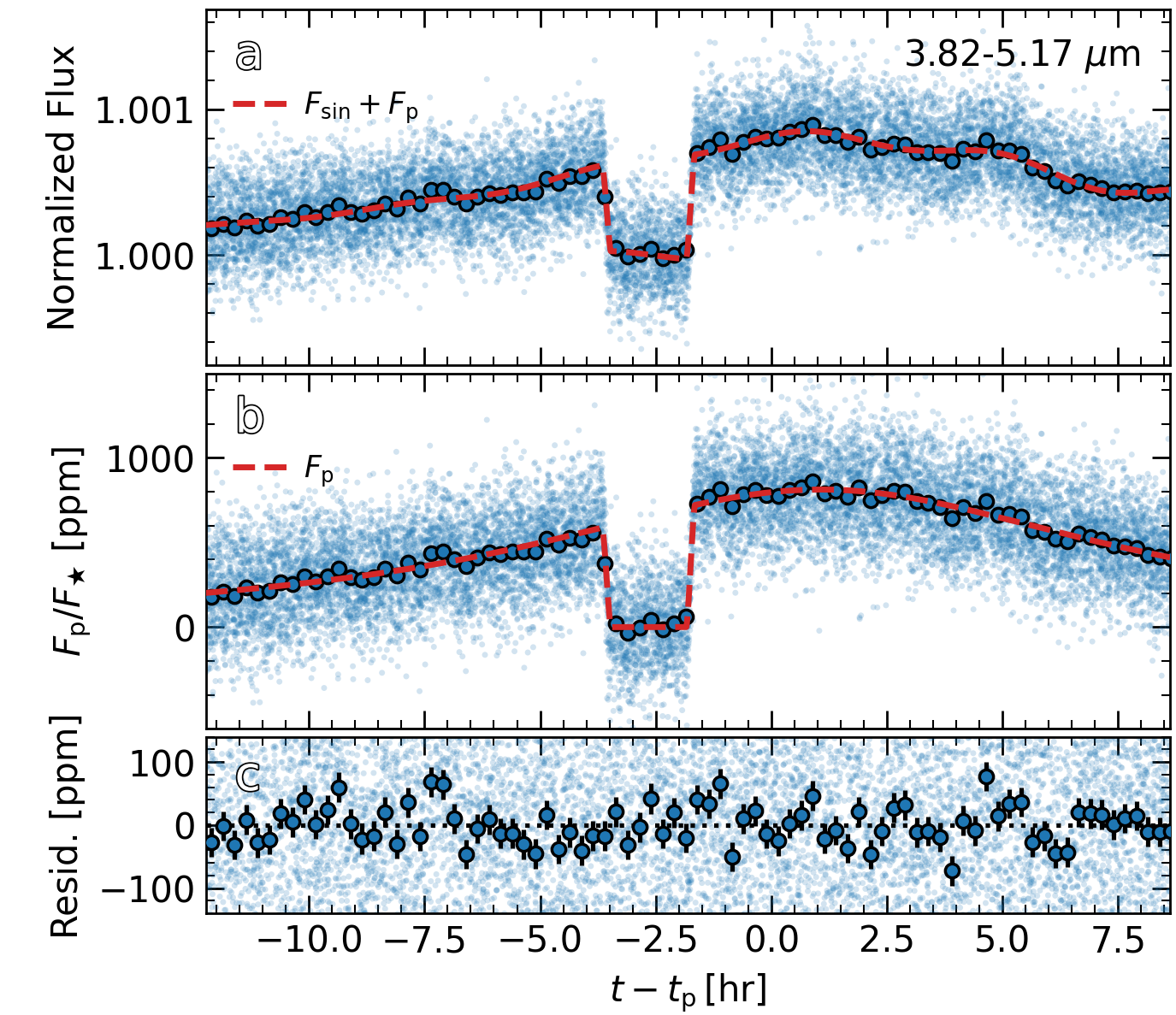}
    \caption{Same as Fig. \ref{fig:NRS2_WL} but for the NRS2 white light curve fit with the nominal model, which consists of a symmetric phase curve, a sinusoidal term ($F_{\rm sin}$) with frequency of $f_1=0.231\,{\rm hr^{-1}}$, and no polynomial term. Panel `a' is the observed flux, in panel `b' $F_{\rm sin}$ has been removed, and panel `c' shows the residuals.}
    \label{fig:NRS2_WL}
\end{figure}

\section{Results}\label{sect:results}

\subsection{White light curves}\label{sect:WLCs}

Each of the integrated NRS1 and NRS2 white light curves obtained from the \texttt{Eureka!} reduction were fitted using the modelling framework described in Sect. \ref{sect:LC_model}. Bayesian Information Criterion (BIC) values were calculated for several variations of the general light curve model in order to select the most statistically significant model with the fewest parameters. Specifically, the goal is to determine whether $F_{\rm poly}$ should be included in the model and what polynomial order is preferred by the data, whether a symmetric or asymmetric planet phase curve is preferred, and whether the $F_{\rm sin}$ and/or $\Delta F_{\rm jump}$ components should be included. The model selection process involves three steps: (1) we fit the data without including an $F_{\rm sin}$ term and using a symmetric/asymmetric $F_{\rm p}$ model and an $F_{\rm poly}$ term up to $3^{\rm rd}$ order; (2) we then test whether including the $\Delta F_{\rm jump}(t)$ term significantly decreases the BIC (where $|\Delta{\rm BIC}|>5$ is considered significant); and (3) whether including the $F_{\rm sin}$ term significantly decreases the BIC.

\subsubsection{NRS1 white light curve}\label{sect:NRS1_WLC}

Based on the first step of the model selection process, we found that the NRS1 white light curve favors a symmetric phase curve model with a $2^{\rm nd}$ order polynomial term, which has a $|\Delta{\rm BIC}|>10$ compared to the models using a $1^{\rm st}$ or $3^{\rm rd}$ order $F_{\rm poly}$ term and symmetric/asymmetric $F_{\rm p}$ model. For steps 2 and 3 of the model selection, we found that including both a $\Delta F_{\rm jump}(t)$ term and an $F_{\rm sin}$ term produced a large decrease in the BIC values: relative to the symmetric $F_{\rm p}$, $2^{\rm nd}$ order $F_{\rm poly}$ model, including $\Delta F_{\rm jump}(t)$ decreases the BIC by $86$ and including $F_{\rm sin}$ decreases this by an additional $244$. Therefore, the nominal model we adopted for the NRS1 white light curve consists of a $2^{\rm nd}$ order $F_{\rm poly}$, symmetric $F_{\rm p}$, and non-zero $\Delta F_{\rm jump}$ and $F_{\rm sin}$ terms and includes 15 free parameters ($F_0$, $c_1$, $c_2$, $c_3$, $T_{\rm e}$, $p_1$, $p_2$, $c_y$, $c_{{\rm GS},y}$, $A_1$, $B_1$, $\phi_1$, $\Delta F_{\rm jump}$, $T_{\rm jump}$, and $\sigma_{\rm Jit}$).

The fit to the \texttt{Eureka!}-reduced NRS1 white light curve using the nominal model is shown in Fig. \ref{fig:NRS1_WL} and the derived parameters are listed in Table \ref{tbl:NRS1_NRS2_WLC}. We find that the model achieves a precision of $2.01\times$ the $81\,{\rm ppm}$ photon noise limit. We also fit the light curve obtained from the custom reduction pipeline (see Sect. \ref{sect:custom_reduction}) for which we obtain a lower precision of $2.44\times$ the photon noise limit. Some red noise is still present in the residuals as seen by eye in Fig. \ref{fig:NRS1_WL} and indicated in the Allen deviation plot shown in Fig. \ref{fig:NRS1_NRS2_SLC_allan} of the Appendix. The measured flux is slightly correlated with $y(t)$ and $y_{{\rm GS},y}(t)$ based on the derived $c_y=-2.6\pm1.4\,{\rm ppm}$ and $c_{{\rm GS},y}=2.5\pm1.4\,{\rm ppm}$. The $1^{\rm st}$ and $2^{\rm nd}$ order polynomial coefficients are significantly non-zero with $p_1=-153.75_{-0.73}^{+0.77}\,{\rm ppm/hr}$ and $p_2=3.396\pm0.078\,{\rm ppm/hr^2}$). As discussed further in Sect. \ref{sect:SLCs} in the context of the spectral light curves as well as in the Appendix, we derive similarly high magnitudes for $p_1$ and $p_2$ from an analysis of the publicly available NIRSpec/G395H transit observations of GJ486b \citep{moran2023}. This suggests that these polynomial trends are not unique to the HD~80606 NIRSpec data and the analysis presented here.

For the $\Delta F_{\rm jump}(t)$ term, we obtain a jump occurrence time of $T_{\rm jump}=2459885.16676_{-0.00044}^{+0.00060}$, which is consistent with the first HGA move, and an amplitude of $\Delta(F_{\rm p}/F_\star)_{\rm jump}=77\pm10\,{\rm ppm}$. We note that the jump correction amplitude is correlated with $c_3$ and $p_1$. As a result, the difference in the inferred planet flux when including the jump term is $\approx77\,{\rm ppm}$ near the eclipse mid-point but is much less further away from this point since the jump in flux is partially absorbed by $c_3$ and $p_2$. Nonetheless, we find that including $\Delta F_{\rm jump}(t)$ does have a non-negligible impact ($\sim4\sigma$) on the derived $c_1$ and $c_3$ values.

\subsubsection{NRS2 white light curve}\label{sect:NRS2_WLC}

For NRS2, the first step of the model selection process yielded the lowest BIC for the model that includes a symmetric $F_{\rm p}$ and does not include a polynomial term: for both the symmetric and asymmetric $F_{\rm p}$ models, including a polynomial increased the BIC by $>7$. For step 2, we find that including the $\Delta F_{\rm jump}(t)$ term led to a higher BIC and, unlike for the NRS1 white light curve, a $T_{\rm jump}$ that was poorly constrained. However, the BIC does decrease significantly by $148$ when including an $F_{\rm sin}$ term for step 3. We therefore adopt a nominal model that includes a symmetric $F_{\rm p}$, an $F_{\rm sin}$ term, and an $F_{\rm sys}$ term with $\Delta F_{\rm jump}=0$, which consists of 11 free parameters ($F_0$, $c_1$, $c_2$, $c_3$, $T_{\rm e}$, $c_y$, $c_{{\rm GS},y}$, $A_1$, $B_1$, $\phi_1$, and $\sigma_{\rm Jit}$).

The fit to the NRS2 white light curve using the nominal model is shown in Fig. \ref{fig:NRS2_WL}. The derived parameters are listed in Table \ref{tbl:NRS1_NRS2_WLC} where we also include for comparison the parameters derived when the jump parameter is not included in the model. The nominal model yields a precision of $2.02\times$ the $108\,{\rm ppm}$ photon noise limit for the \texttt{Eureka!} reduction and $2.45\times$ the photon noise for the light curve obtained from the custom reduction. As seen in the Allen deviation plot in the Appendix (Fig. \ref{fig:NRS1_NRS2_SLC_allan}), the NRS2 residuals have red noise with a timescale comparable to that found in the NRS1 residuals.

Comparing the NRS2 and NRS1 phase curves, we find large differences with the former having a much higher amplitude ($c_1$), peak offset ($c_2)$, and rise/decay timescale ($c_3$), which all differ by $\gtrsim5\sigma$. The eclipse mid-point times differ by about $4.3\sigma$ (${\rm BJD}=2459885.16528\pm0.00025$ and ${\rm BJD}=2459885.16420\pm0.00016$ for NRS1 and NRS2, respectively). This difference may be significant, however, when the jump parameter is not included in the NRS1 model, we derive a lower $T_{\rm e}$ that has a $\sim2.4\sigma$ difference with respect to NRS2. \citet{pearson2022} derive an eclipse mid-point of $T_{\rm e}^\prime=2458882.214\pm0.0021\,{\rm BJD}$. Using the reported $P_{\rm orb}=111.436765\pm0.000074\,{\rm d}$, we find a discrepancy between the predicted eclipse time and the value derived from the NRS2 white light curve of $T_{\rm e}^\prime-T_{\rm e}^{\rm NRS2}=-28\pm4\,{\rm min}$. This discrepancy may be partly attributed to an underestimated uncertainty in the published $T_{\rm e}^\prime$, which \citet{pearson2022} note was derived using a fixed $\omega$ value ($\omega=-58.887\pm0.043\,{\rm deg}$).

The NRS2 bandpass is comparable to the \emph{Spitzer} $4.5\,{\rm \mu m}$ bandpass allowing for rough comparisons to be made with the phase curve presented by \citet{wit2016}. These authors report an amplitude of $738\pm52\,{\rm ppm}$ and is therefore comparable to our value of $812.9\pm6.8\,{\rm ppm}$. The \emph{Spitzer} $4.5\,{\rm \mu m}$ phase curve has a notably lower peak offset of between $\approx-1\,{\rm hr}$ and $0\,{\rm hr}$ relative to periapse (for NRS2, we derive an offset of $1.073\pm0.040\,{\rm hr}$). The timescale over which the planet flux decreases post-periapse is also noticeably higher for NRS2 relative to the \emph{Spitzer} phase curve: between the time of peak flux and $t-t_{\rm p}\approx8\,{\rm hr}$ (near the end of our observing window), $F_{\rm p}/F_\star$ decreases from $\approx830\,{\rm ppm}$ to $450\,{\rm ppm}$ compared to a decrease of $\approx740\,{\rm ppm}$ to $180\,{\rm ppm}$ for \emph{Spitzer}. On the other hand, the rise in flux from $t-t_{\rm p}\approx-12\,{\rm hr}$ associated with the two phase curves are in close agreement.

\subsection{Spectral light curves}\label{sect:SLCs}

\begin{figure}
	\centering
    \includegraphics[width=1\columnwidth]{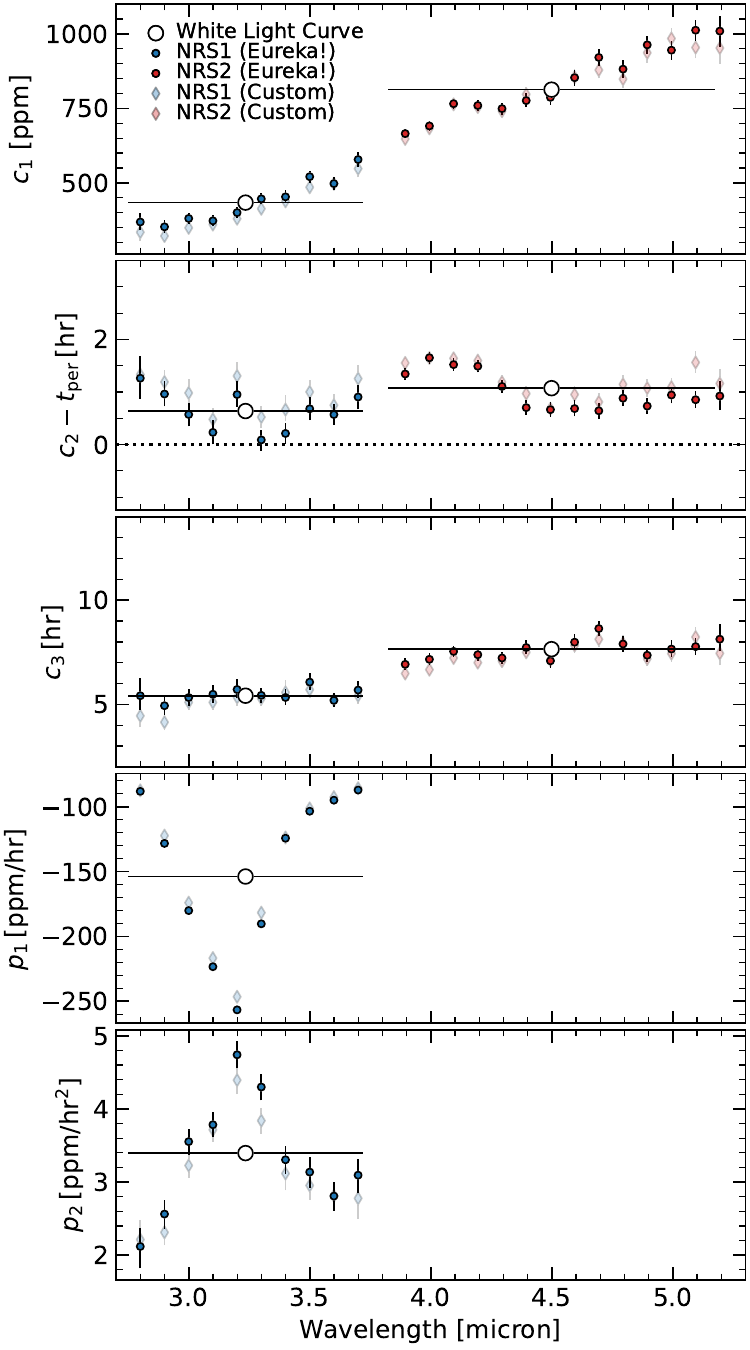}
	\caption{Phase curve amplitude ($c_1$), peak offset ($c_2$), and rise/decay timescale ($c_3$) along with the polynomial coefficients ($p_1$ and $p_2$). Darker blue and red circles correspond to the NRS1 and NRS2 spectral light curves reduced using \texttt{Eureka!}. White circles are the NRS1/NRS2 white light curve values where the horizontal error bars indicate the wavelength bin widths. Lighter diamonds are obtained using the custom reduction pipeline. Note that the nominal NRS2 model does not include a polynomial term, which is why there are no associated $p_1$ and $p_2$ terms plotted in the $4^{\rm th}$ and $5^{\rm th}$ panels.}
	\label{fig:ph_curve_param}
\end{figure}

Each of the \texttt{Eureka!} spectral light curves (10 for NRS1 and 14 for NRS2) was fitted using the nominal models adopted for the white light curves (Sect. \ref{sect:WLCs}). For comparison, we also fit the spectral light curves obtained from the custom data reduction pipeline (Sect. \ref{sect:custom_reduction}). The derived phase curve parameters associated with each of the NRS1 and NRS2 spectral light curves (and those derived for the white light curves) are plotted in Fig. \ref{fig:ph_curve_param} along with the polynomial coefficients included in the nominal NRS1 model. In general, we find that the two reductions yield comparable results, which exhibit the same trends but have some systematic discrepancies. Compared to the custom reduction, the \texttt{Eureka!} NRS1 reduction shows $\sim2\sigma$ higher $c_1$ and $p_2$ values and $\sim2\sigma$ lower $c_2$ and $p_1$ values, particularly near $3.2\,{\rm \mu m}$. These differences are partly explained by the positive/negative correlation between $c_1$/$c_2$ and $p_2$ and the relatively weak negative correlation between $c_2$ and $p_1$ (see the corner plot in Fig. \ref{fig:nrs1_wlc_corner} in the Appendix). The NRS2 light curves generally show closer agreement with typical discrepancies $\lesssim1\sigma$.

As expected based on the observed light curves shown in Fig. \ref{fig:NRS1_NRS2_2D}, the linear slope ($p_1$ in Eqn. \ref{eqn:Fpoly}) included in the NRS1 model's polynomial term is highly correlated with wavelength. The distribution of $p_1(\lambda)$ is roughly Gaussian in shape with a maximum magnitude occurring at $\approx3.2\,{\rm \mu m}$. The second-order polynomial coefficient ($p_2$) shows a characteristically similar wavelength-dependence with a peak occurring at the same wavelength. The highly wavelength-dependent systematics affecting our NRS1 observations are also shown to impact NIRSpec/G395H transit observations of GJ486b \citet{moran2023} (see their Fig. 1). Both data sets were obtained using a similar instrument configuration using low group numbers in order to avoid saturation (HD~80606 was observed using 5 groups while those of GJ486 used 3). We analyzed the out-of-transit baseline exposures from the publicly available GJ486 data in order to characterize the wavelength dependence of $p_1$ and $p_2$ (see the Appendix). This analysis yields a similar Gaussian-like trend in $p_1(\lambda)$ while $p_2$ is nearly constant for $\lambda\gtrsim3.1\,{\rm \mu m}$ with a value of $\approx60\,{\rm ppm/hr^2}$ (Fig. \ref{fig:gj486b_p1p2}).

The phase curve amplitudes ($c_1$) show a nearly monotonic increase with wavelength from $\sim400\,{\rm ppm}$ at $2.8\,{\rm \mu m}$ to $\sim1000\,{\rm ppm}$ at $5.2\,{\rm \mu m}$. Small decreases in $c_1$ are apparent at $\approx3.0-3.5\,{\rm \mu m}$ and at $\approx4.2-5.1\,{\rm \mu m}$. Similar decreases are apparent in the peak offset ($c_2$), which range from $\approx0-1.3\,{\rm hr}$ within the NRS1 bandpass and from $\approx0.5-1.5\,{\rm hr}$ within NRS2. A sharp increase in $c_2$ coinciding with the minimum $p_1$ value and the maximum $p_2$ value near $3.2\,{\rm\mu m}$ is apparent. As noted above, the $c_2$ parameter is negatively correlated with $p_2$ and, to a lesser extent, $p_1$; therefore some of the increase in $c_2$ may be due to the decrease in $p_1$. Lastly, the rise/decay timescale ($c_3$) shows a gradual increase across both the NRS1 and NRS2 wavelengths ranging from $\approx5-9\,{\rm hr}$ along with small decreases most notably at $4.5\,{\rm \mu m}$ and $4.9-5.2\,{\rm \mu m}$. We note that, while the polynomial terms included in the NRS1 model may somewhat bias the inferred phase curve parameters, the NRS2 light curves, which do not have obvious wavelength-correlated noise, show comparable trends.

\begin{figure}
	\centering
     \includegraphics[width=1\columnwidth]{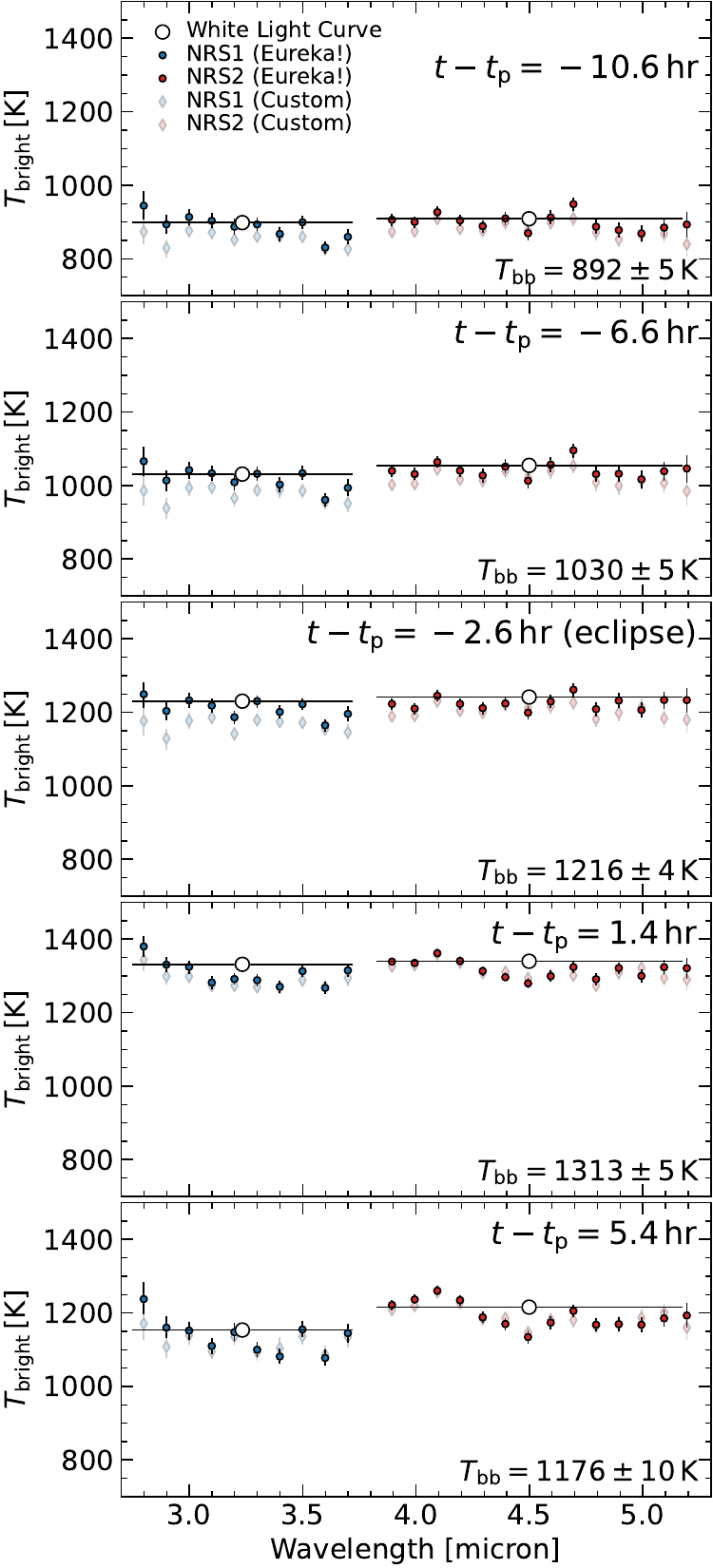}
	\caption{Brightness temperatures derived using the nominal NRS1 and NRS2 light curve models for the \emph{Eureka!} reduction (dark circles) and for the custom reduction pipeline (light diamonds). The estimated blackbody temperatures ($T_{\rm bb}$) derived by fitting a blackbody spectrum to the NRS1 and NRS2 $F_{\rm p}/F_\star$ measurements while including a flux offset for NRS1 (see Sect. \ref{sect:SLC_spec}) are shown for each phase. White circles correspond to values calculated from the white light curves.}
	\label{fig:Tb_spec}
\end{figure}

\subsection{Phase-resolved planet spectra}\label{sect:SLC_spec}

We used the fitted spectral light curves to extract the planet-to-star flux contrast ($F_{\rm p}/F_\star$) and the planet's brightness temperatures ($T_{\rm bright}$) at several phases in order to identify and characterize any time variability in the planet's emission spectrum. Eclipse depths were calculated from the mean $F_{\rm p}/F_\star$ values associated with the phase curve model averaged over a $2\,{\rm hr}$ window centered on the eclipse mid-point. We then carried this out for nine additional phases totaling four pre-eclipse phases and six post-eclipse phases. The $F_{\rm p}/F_\star$ spectra are presented in Sect. \ref{sect:retrieval_results} alongside the atmospheric retrieval results. For reference, all ten spectra are shown in Fig. \ref{fig:FpFs_spec} in the Appendix where we include both the nominal spectra obtained from the \texttt{Eureka!} reduction and the spectra obtained from the custom reduction.

Brightness temperatures were calculated for each wavelength and orbital phase using the $F_{\rm p}/F_\star$ measurements. The stellar spectrum used to derive the planet's thermal emission ($F_{\rm p}$) was estimated using a PHOENIX model with $T_{\rm eff}=5600\,{\rm K}$, $\log{g}=4.5$, and ${\rm [M/H]}=0.5$. Instrumental broadening assuming $R=2700$ was applied to the model stellar spectrum, which was then interpolated onto the native NIRSpec resolution and binned using the $0.1\,{\rm \mu m}$ bin widths adopted for the observed spectral light curves. The planet-to-star radius ratio used for the calculation of $T_{\rm bright}$ was taken to be $(R_{\rm p}/R_\star)^2=0.01019\pm0.00023$ \citep{pearson2022}. The brightness temperatures derived for five of the ten phases spanning the observing window are shown in Fig. \ref{fig:Tb_spec}. Comparing $T_{\rm bright}$ calculated from both the \texttt{Eureka!} and the custom reduction pipelines shows generally good agreement. During the pre-periapse phases, the \texttt{Eureka!} reduction yields $T_{\rm bright}$ values within the NRS1 bandpass that are systematically higher than the custom reduction by $\sim50\,{\rm K}$ while the NRS2 wavelengths agree within $\lesssim1\sigma$. The NRS1 discrepancy may be attributed to the correlation between the $c_1$/$c_2$ phase curve parameters and $p_2$ noted above in Sect. \ref{sect:SLCs}.

Next, we proceeded to fit the $F_{\rm p}/F_\star$ spectra for each phase assuming that the planet radiates as a blackbody. Best-fitting blackbody temperatures and uncertainties were estimated by minimizing the $\chi^2$ values using the \texttt{curve\_fit} function from the \texttt{scipy} Python package where the stellar flux is given by the PHOENIX stellar model noted above. Fits were carried out using the \texttt{Eureka!}-reduced light curves for three detector configurations: using both NRS1 and NRS2 measurements, using both NRS1 and NRS2 measurements while also including an additive flux free parameter to the NRS1 measurements ($f_{\rm NRS1}$) in order to account for possible offsets related to the instrument or introduced by the NRS1 light curve polynomial detrending term, and using only the NRS2 measurements. In all three cases, we find a similar trend in which the first six phases have relatively low reduced $\chi^2$ values ($0.6<\chi^2_{\rm red}<1.7$) compared to the last four phases ($2.2<\chi^2_{\rm red}<7.0$). Comparing the derived $T_{\rm bright}$ values for the NRS1+NRS2 and NRS1+NRS2+$f_{\rm NRS1}$ cases, we find discrepancies of $0.5-2.6\sigma$ with the last three phases showing the largest differences. Slightly larger discrepancies are found when comparing the NRS1+NRS2 and NRS2-only configurations ($0.8-2.6\sigma$) while the NRS1+NRS2+$f_{\rm NRS1}$ and NRS2-only cases show negligible differences ($\lesssim0.5\sigma$). The $T_{\rm bright}$ values for case 1 are shown in Fig. \ref{fig:Tb_spec}, which increase from $892\pm5\,{\rm K}$ at the start of the observing window to a peak value of $1313\pm5\,{\rm K}$ during the first post-periapse phase ($t-t_{\rm p}=1.4\,{\rm hr}$) before decreasing to $1077\pm10\,{\rm K}$.

In Fig. \ref{fig:Tb_spec}, the post-periapse phases show the largest deviations from a blackbody with the appearance of several potential absorption features. Within the NRS2 bandpass, a decrease in $T_{\rm bright}$ is found at $\lambda>4.3\,{\rm \mu m}$, which is qualitatively consistent with CO and CO$_2$ absorption \citep[cf. recent NIRCam eclipse measurements of HD 149026b,][]{bean2023}. Within the NRS1 bandpass, a decrease in $T_{\rm bright}$ is found from $3.1-3.4\,{\rm \mu m}$, which may be due to absorption by CH$_4$ \citep[cf. NIRCam eclipse measurements of WASP-80b,][]{bell2023}. These features are discussed further below.

\begin{figure*}
	\centering
    \includegraphics[height=0.36\textwidth]{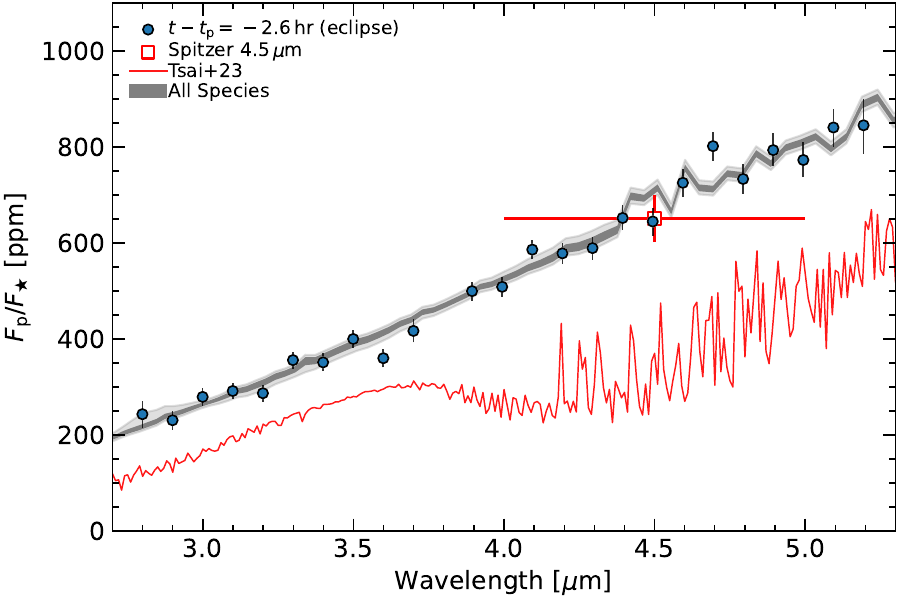}\vspace{-0.05cm}
    \includegraphics[height=0.36\textwidth]{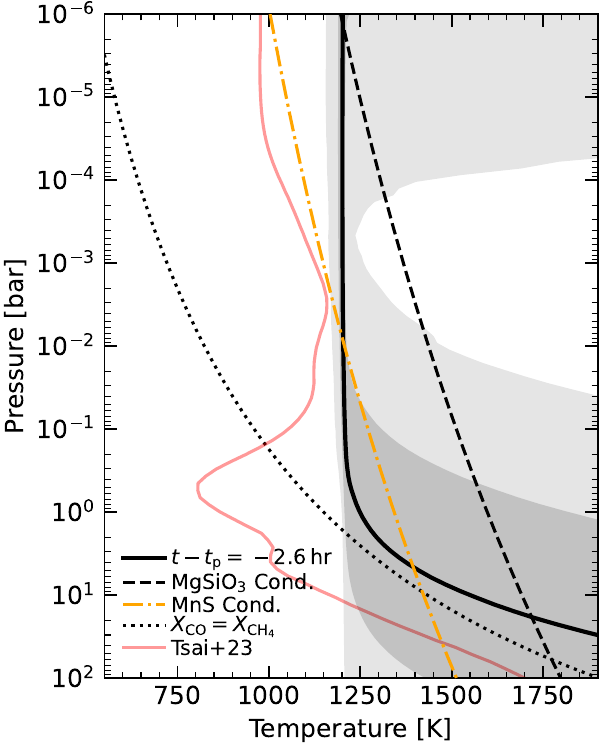}\vspace{-0.05cm}

    \includegraphics[height=0.36\textwidth]{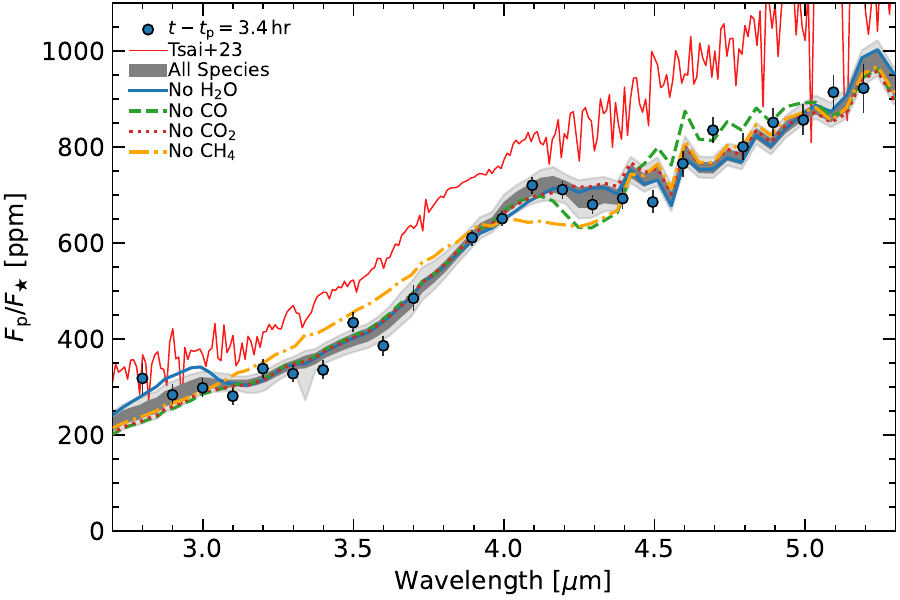}\vspace{-0.05cm}
    \includegraphics[height=0.36\textwidth]{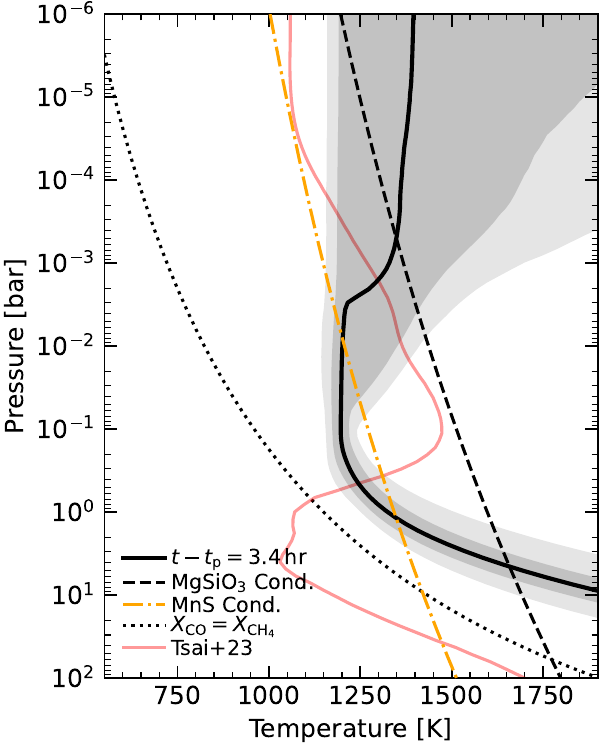}\vspace{-0.05cm}
    \caption{Results obtained from the \texttt{petitRADTRANS}-based chemical equilibrium retrievals for two phases: at the eclipse (top row), which is consistent with a blackbody with a $1219\pm23\,{\rm K}$ temperature, and at the $8^{\rm th}$ extracted phase (bottom row), which shows the maximum CH$_4$ detection significance. \emph{Left column:} the observed NRS1 ($\lambda<3.8\,{\rm \mu m}$) and NRS2 ($\lambda>3.8\,{\rm \mu m}$) emission spectra (blue circles) compared with the $1\sigma$ (dark gray) and $2\sigma$ (light gray) uncertainty bands from the retrievals. The MAP chemical equilibrium solutions obtained when removing one of the four included gasseous opacities, which were calculated to evaluate the detection significance, are plotted for the bottom row. The red square is the $4.5\,{\rm \mu m}$ \emph{Spitzer} eclipse depth \citep{wit2016}. \emph{Right column:} PT profiles derived for the retrievals where the solid black line is the median profile and the dark/light bands indicate $1\sigma/2\sigma$ uncertainties. The faint red lines are the GCM-based profiles published by \citet{tsai2023} assuming a solar metallicity and $T_{\rm int}=100\,{\rm K}$. The dotted line indicates points at which CO and CH$_4$ are in equal abundance for a solar metallicity atmosphere in chemical equilibrium \citep{fortney2020} while the yellow dash-dotted and black dashed lines are the condensation curves for MnS and MgSiO$_3$ \citep{morley2012,visscher2010}.}
	\label{fig:FpFs_PT}
\end{figure*}

\begin{figure}
	\centering
    \includegraphics[width=1\columnwidth]{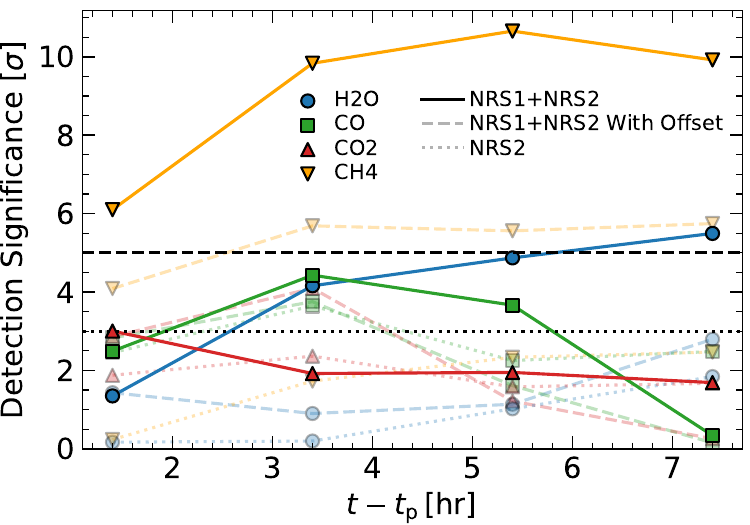}
	\caption{Detection significance levels estimated for each molecule at each of the five post-periapse phases (where the $x$-axis corresponds to time relative to periapse). Dark symbols/solid lines are obtained using both NRS1 and NRS2, light symbols/dashed lines use NRS1 and NRS2 with a flux offset, and light symbols/dotted lines use only NRS2.}
	\label{fig:detect_sig}
\end{figure}

\begin{figure}
	\centering
    \hspace{-0.48cm}
    \includegraphics[width=1.0\columnwidth]{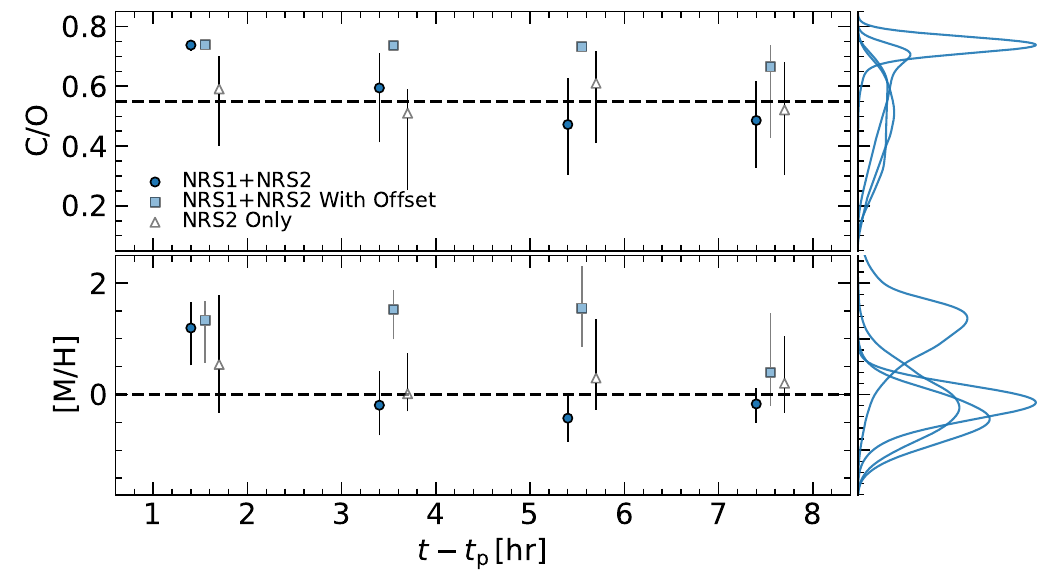}
    \caption{Atmospheric compositions inferred from the chemical equilibrium retrievals for the four post-peripase phases that deviate from a blackbody spectrum. We show the C/O ratio (top panel) and the metallicity (${\rm [M/H]}$) (bottom panel) for case 1 (using NRS1 and NRS2; dark circles), case 2 (using NRS1 and NRS2 and a flux offset; light squares), and case 3 (using only NRS2; white triangles). The horizontal dashed lines indicate solar values. The marginalized posterior distributions associated with case 1 are plotted along the right axis.}
	\label{fig:composition}
\end{figure}

\subsection{Atmospheric retrievals}\label{sect:retrieval_results}

We carried out the \texttt{petitRADTRANS}-based chemical equilibrium atmospheric retrievals described in Sect. \ref{sect:retrievals} for the three instrument configurations that we considered in conjunction with the blackbody model (Sect. \ref{sect:SLC_spec}): the NRS1+NRS2 case, the NRS1+NRS2+$f_{\rm NRS1}$ case, and the NRS2-only case. BIC values were then calculated for the maximum a posteriori (MAP) solutions obtained from these retrievals and were compared with the blackbody spectrum MAP solution BIC values. We find that for both the NRS1+NRS2 and NRS1+NRS2+$f_{\rm NRS1}$ cases, the chemical equilibrium model is preferred over a simple blackbody spectrum for the four post-periapse phases. This is based on the calculated $\Delta{\rm BIC}$ values ranging from $-6$ to $-112$ where the NRS1+NRS2 case exhibits the largest differences ($<-16$); for the NRS2-only case, the chemical equilibrium model is preferred only for the last three phases ($-25<\Delta{\rm BIC}<-16$). For each of the four phases in which we find significant deviations from a blackbody spectrum, we carried out a similar BIC comparison between the \texttt{petitRADTRANS}-based chemical equilibrium retrievals and the \texttt{PyratBay}-based free retrievals. Both modeling frameworks yield comparable $\chi^2$ values and the chemical equilibrium model, having fewer free parameters, is favored over the free retrievals. Therefore, based on the quality of fits provided by each considered model, we conclude that during the first six phases, the observed spectrum is indistinguishable from a blackbody spectrum; the last four phases are best characterized by a chemical equilibrium model atmosphere and do not show evidence of disequilibrium chemistry or photochemistry.

In Fig. \ref{fig:FpFs_PT}, we show the results of the chemical equilibrium retrievals carried out for the NRS1+NRS2 case at the eclipse phase ($t-t_{\rm p}=-2.6\,{\rm hr}$) and at the second post-periapse phase ($t-t_{\rm p}=3.4\,{\rm hr}$). The left columns compare the observed spectra with models generated from the retrieval posteriors and with the GCM-based predictions published by \citet{tsai2023}. The GCM-based predictions assume a solar-metallicity atmosphere with a $100\,{\rm K}$ internal temperature. The right column shows the derived PT profile constraints along with the GCM PT profiles. All four of the post-periapse spectra and PT profiles are also shown in Fig. \ref{fig:FpFs_PT_all} of the Appendix. We find that the PT profiles derived from the chemical equilibrium retrievals for all three detector combinations are consistent in terms of the overall profile shape and in terms of their photospheric temperatures at pressures of $1-10\,{\rm mbar}$. Similar agreement is found when comparing the profiles derived from the chemical equilibrium and \texttt{PyratBay}-based free retrievals for the NRS1+NRS2 case. During the first five phases through the eclipse ($-10.6\leq t-t_{\rm p}\leq-2.6\,{\rm hr}$), the derived blackbody temperatures and photospheric temperatures associated with the retrievals are notably higher than those of the GCM PT profiles. Phases six and seven ($-0.6\leq t-t_{\rm p}\leq1.4\,{\rm hr}$) show good agreement while the GCM temperatures are higher during the last three phases. These differences are manifest in the spectra as a vertical shift across the observed bandpasses (see Fig. \ref{fig:FpFs_spec} in the Appendix). Additionally, none of the retrieved PT profiles show clear evidence of an inversion near the infrared photosphere as found in the GCMs and no obvious emission features are seen in the extracted NIRSpec spectra that would otherwise indicate an inversion.

The posteriors for the pressure of the gray cloud deck included in the retrievals are poorly constrained during the first six phases. For the post-eclipse phases, we obtain $2\sigma$ lower limits on the cloud deck's altitude at pressure layers $P_{\rm cloud}\gtrsim-0.8\,{\rm bar}$ for the chemical equilibrium retrievals and $P_{\rm cloud}\gtrsim-1.1\,{\rm bar}$ for the free retrievals (i.e., below the $1-10\,{\rm mbar}$ IR photosphere). As shown in Fig. \ref{fig:FpFs_PT}, the increase in temperature occurring from the pre- to post-periapse phases is such that the photospheric temperatures reach the condensation point of MnS clouds (yellow dash-dotted line) \citep{morley2012} suggesting that such clouds, if present, are expected to dissipate during the $21\,{\rm hr}$ observing window (discussed further in Sect. \ref{sect:disc}).

We estimated the detection significance of H$_2$O, CO, CO$_2$, and CH$_4$ for the last four extracted phases using a similar method presented by \citet{bell2022}. This involves carrying out additional chemical equilibrium retrievals using \texttt{petitRADTRANS} with each of the four gas species being excluded from the model one at a time. We used \texttt{emcee} to generate 16 chains $2\times10^4$ steps in length each with the first $10^4$ steps discarded as burn-in yielding $1.6\times10^5$ samples. From these samples we then selected the maximum likelihood solution and compared the associated $\chi^2$ value with that obtained when including all gas species. The resulting significance levels for each molecule and for each phase are shown in Fig. \ref{fig:detect_sig}. Adopting a $3\sigma$ threshold, we find that the NRS1+NRS2 case yields detections of CH$_4$ ($6.1-10.7\sigma$), H$_2$O ($4.2-5.5\sigma$), and CO ($3.7-4.4\sigma$). CH$_4$, CO, and CO$_2$ have $>3\sigma$ significance levels for the NRS1+NRS2+$f_{\rm NRS1}$ case ($4.1-5.7\sigma$, $3.8\sigma$, and $4.1\sigma$ respectively) while only CO is detected for the NRS2-only case ($3.6\sigma$).

The MAP solutions found for the 4 models without H$_2$O, CO, CO$_2$, or CH$_4$ are overplotted in Fig. \ref{fig:FpFs_PT} and in Fig. \ref{fig:FpFs_PT_all} of the Appendix; comparing these with the solutions obtained using all species demonstrates which wavelength channels have the largest contribution to the estimated detection significance. Both the CO and CO$_2$ absorption bands are primarily found within the NRS2 bandpass and therefore should be detectable when using only NRS2. Since CO$_2$ is not detected in this case and is only detected during a single phase for the NRS1+NRS2+$f_{\rm NRS1}$ case, the detection is not considered to be reliable.

In Fig. \ref{fig:composition}, the C/O ratio and metallicity (${\rm [M/H]}$) derived from the \texttt{petitRADTRANS} chemical equilibrium retrievals are plotted for the last four phases using all three instrument configurations. The NRS2-only configuration exhibits solar C/O and ${\rm [M/H]}$ values for each phase and has the largest uncertainties due to the exclusion of the NRS1 bandpass but is also not impacted by potential flux offsets. During the last three phases, these C/O and ${\rm [M/H]}$ values are in agreement with the NRS1+NRS2 case where the highest precision is obtained for the last phase (${\rm C/O}=0.49\pm0.15$ and ${\rm [M/H]}=-0.17\pm0.31$). Super-solar C/O and ${\rm [M/H]}$ values are obtained for the first post-periapse phase in the NRS1+NRS2 case and for the first to third post-peripase phases when including $f_{\rm NRS1}$. These instances are all associated with non-detections of H$_2$O, which therefore likely biases the inferred C/O and ${\rm [M/H]}$ since, within the \texttt{petitRADTRANS} chemical equilibrium model framework, C/O is controlled by varying the total O abundance.

Considering the overall consistency between the C/O and ${\rm [M/H]}$ values derived for the NRS1+NRS2 and NRS2-only configurations, we conclude that the NRS1+NRS2 case provides the most precise and reliable constraints on the planet's atmosphere during the last three observed phases where we obtain $>3.6\sigma$ detections for H$_2$O, CO, and CH$_4$ during at least two of these phases.

\section{Discussion}\label{sect:disc}

\subsection{Partial phase curves}

The analysis presented here demonstrates both the feasibility and challenges of using partial phase curve observations obtained with NIRSpec/G395H to study hot Jupiter atmospheres. The long-wavelength NRS2 detector ($3.8-5.2\,{\rm \mu m}$), which is largely sensitive to CO, is not found to exhibit obvious systematics that may potentially bias the inferred planetary signal. On the other hand, the short-wavelength NRS1 detector ($2.8-3.7\,{\rm \mu m}$), which is largely sensitive to CH$_4$ and H$_2$O, does have systematics that manifest as strong, non-linear and wavelength-dependent slopes (Fig. \ref{fig:NRS1_NRS2_2D}). Based on the derived polynomial coefficients, these slopes are consistent for the two independent data reduction pipelines used in this work (see Fig. \ref{fig:ph_curve_param}) suggesting that the trends cannot easily be removed at the reduction level and/or must be detrended using additional parameters not considered in our light curve model. The small number of NIRSpec/G395H time series data sets that are publicly available suggest that bright targets observed with few group numbers like HD~80606 ($J=7.7\,{\rm mag}$) and GJ486 ($J=7.2\,{\rm mag}$) may be more strongly impacted by these systematics compared to dimmer targets observed with much higher group numbers such as WASP-39 ($J=10.7\,{\rm mag}$) \citep[e.g.,][]{alderson2023,moran2023}; therefore, partial phase curves of such bright targets using the NIRSpec/G395H instrument mode should be approached with caution until a robust method of removing the systematic trends is developed.

Comparing the phase curve parameters derived from the NRS1 and NRS2 detectors suggests that our adopted method of detrending the NRS1 systematic slopes in order to accurately recover the planetary signal is effective but likely imperfect. For instance, while the phase curve amplitudes ($c_1$) and peak offsets ($c_2$) do not exhibit an obvious jump between the two detectors, the NRS1 rise/decay timescale ($c_3$) potentially show a systematic offset (Fig. \ref{fig:ph_curve_param}). The $c_3$ parameter is correlated with the polynomial's $1^{\rm st}$ order coefficient ($p_1$) such that a steeper inferred slope may yield a higher $c_3$. 

We note that the $2^{\rm nd}$ order polynomial used to detrend the NRS1 systematics would likely be more difficult and more susceptible to biases if the planet phase curve is not symmetric (our analyses of the NRS1 and NRS2 light curves independently indicate a highly symmetric phase curve). This is apparent from several of the NRS1 spectral light curves in which, when fitting using the asymmetric phase curve model, we obtain a much higher $c_4$ value compared to the $c_3$ value. This is not the case for the NRS1 white light curve, which yields consistent $c_3$ and $c_4$ values (thus, the asymmetric model was rejected in favour of the symmetric model).

\begin{figure}
	\centering
    \includegraphics[width=1\columnwidth]{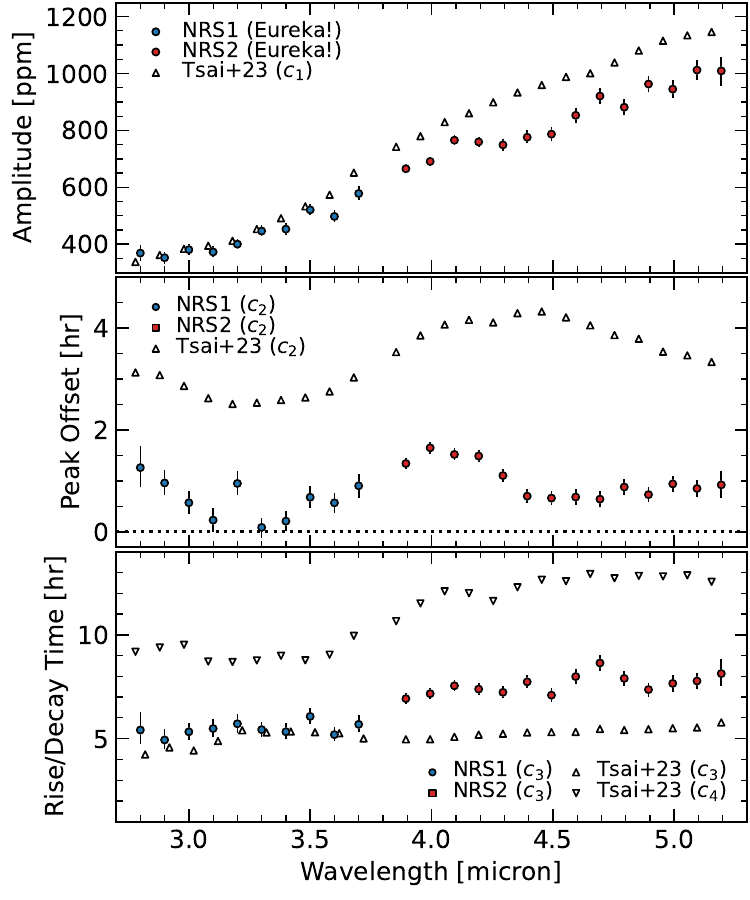}
	\caption{Phase curve amplitude ($c_1$, top panel), peak offset measured with respect to periapse ($c_2-t_{\rm per}$, middle panel), and rise/decay timescale ($c_3/c_4$, bottom panel) derived from the NRS1 and NRS2 data using the nominal (symmetric) model (blue circles and red squares, respectively) compared with those derived from the GCMs published by \citet{tsai2023} (white triangles). The GCM phase curves are highly asymmetric and were therefore fit using the asymmetric model that includes both $c_3$ (the rise timescale) and $c_4$ (the decay timescale).}
	\label{fig:tsai_ph_curve}
\end{figure}

\subsection{Phase curve properties and clouds}

In Fig. \ref{fig:tsai_ph_curve}, we compare the phase curve parameters derived from the NRS1 and NRS2 spectral light curves with those obtained by fitting the GCM phase curves computed by \citet{tsai2023}. The GCM phase curves were fit using the asymmetric version of the analytic phase curve model (Eqn. \ref{eqn:Fp}) where $c_1$, $c_3$, and $c_4$ were included as free parameters and $c_2$ was fixed at the time of peak flux. We find that the phase curve amplitudes are generally in good agreement particularly in the case of NRS1, which show good agreement between $2.7-3.5\,\mu{\rm m}$ before decreasing by $\sim50\,{\rm ppm}$ at $\lambda>3.5\mu{\rm m}$. The measured NRS2 amplitudes are slightly lower than the GCM predictions, which may be partially attributed to the CO absorption features at $\lambda>4.3\,{\rm \mu m}$ seen in our post-periapse spectra. The observed peak offsets, which are expected to be sensitive to the planet's rotation period/wind speed (e.g., Fig. 13 of \citealt{cowan2011} and Fig. 8 of \citealt{lewis2013}), are significantly lower than the GCMs ($\sim0-1.5\,{\rm hr}$ compared to $\sim2.5-4.5\,{\rm hr}$). The narrow drop in the offset near $3.3\,{\rm \mu m}$, which is seen in both the models and in the observations, is potentially attributed to methane in the atmosphere based on the fact that it mirrors the characteristic spectroscopic methane feature (e.g., see the GCM model plotted in Fig. \ref{fig:FpFs_PT}). A similar decrease in the observed peak offset is also apparent within the CO absorption band at $\lambda\gtrsim4.3\,{\rm \mu m}$. In both cases, these peak offset decreases may be caused by the fact that the molecular absorption bands probe higher in the atmosphere, a feature also found in \emph{HST} and \emph{Spitzer} phase curve observations of WASP-43b \citep[e.g.,][]{stevenson2014a,stevenson2017}.

\begin{figure}
	\centering
    \includegraphics[width=1\columnwidth]{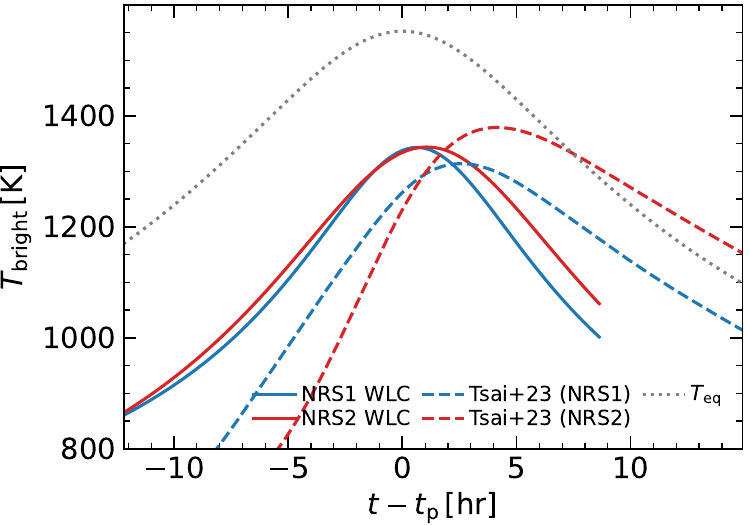}
	\caption{Brightness temperatures calculated from the best-fitting NRS1 and NRS2 phase curves (solid blue and red, respectively) compared with phase curves calculated using the GCMs published by \citet{tsai2023} over the same NRS1/NRS2 bandpasses. The dotted black line shows the instantaneous irradiation temperature calculated assuming full heat redistribution and zero albedo.}
	\label{fig:Tbright_ph}
\end{figure}

Unlike for the observed NRS1 and NRS2 phase curves, the GCM phase curves are highly asymmetric and exhibit much longer decay timescales. This is evident in Fig. \ref{fig:Tbright_ph}, which shows brightness temperatures associated with the integrated NRS1 and NRS2 best-fitting phase curves. A shorter observed rise/decay timescale may indicate that HD~80606b has a shorter radiative timescale and/or a shorter rotation period than that associated with the GCM predictions published by \citet{tsai2023}. A rotation period that is shorter than the assumed $\approx40\,{\rm hr}$ pseudo-synchronous rotation period \citep{hut1981} (i.e., a higher apparent wind speed) will result in more of the planet's cooler side becoming visible towards the end of the observing window causing a more rapid cooling rate to be inferred.

The more rapid decrease in flux relative to the \citet{tsai2023} cloud-free GCM predictions may also be due to the advection of nightside clouds onto the dayside during the post-periapse phases, which would cause the observed flux to be suppressed. This scenario was proposed by \citet{wit2016} and \citet{lewis2017} to explain the low brightness temperatures derived from HD~80606b's \emph{Spitzer} $4.5\,{\rm \mu m}$ phase curve. As noted in Sect. \ref{sect:NRS2_WLC}, the rate at which the \emph{Spitzer} $4.5\,{\rm \mu m}$ flux rises during the pre-periapse phase is consistent with the NRS2 white light curve while the post-periapse decrease in flux is notably slower compared to NRS2. This could be caused by differences in the distribution of the advected clouds between the two observing epochs.

Based on our analysis, HD~80606b's emission spectrum appears to transition from one that is initially indistinguishable from a blackbody to one that exhibits detectable molecular absorption features of H$_2$O, CO, and CH$_4$. It is plausible that this transition is caused by the rapid evaporation of clouds near the photosphere, which may block flux emitted from below the cloud deck. This scenario is consistent with the fact that the transition occurs when the photospheric temperature near $10\,{\rm mbar}$ increases to $\sim1200\,{\rm K}$ coinciding with the condensation point of MnS clouds \citep{morley2012}. Along with MgSiO$_3$ and Na$_2$S, MnS was previously identified as a potentially significant condensate in HD~80606b's atmosphere using GCMs post-processed to include clouds \citep{lewis2017}.

\subsection{Temperature structure}

Ignoring the potential impact of clouds on the observed emission spectral features, the transition from a blackbody spectrum pre-periapse may simply be caused by a change in the temperature gradient near the photosphere. In this case, a (nearly) isothermal PT profile will produce weak or non-existent absorption/emission features. We find that the derived PT profiles at all phases exhibit relatively low temperature gradients near the photosphere and we do not find clear evidence of a temperature inversion near the IR photosphere. This stems from the lack of obvious emission features appearing in any of the extracted emission spectra. The GCM-based predictions from \citet{tsai2023} show a transient inversion forming near the IR photosphere that persists throughout the $21\,{\rm hr}$ observing window, which leads to the formation of subtle but detectable emission features (e.g., Fig. \ref{fig:FpFs_spec}). \citet{iro2010}, \citet{lewis2017}, and \citet{mayorga2021} make similar predictions based on 1D radiative transfer models, 3D GCMs, and 1D radiative-convective equilibrium models, respectively.

\subsection{Chemical composition}

The atmospheric retrievals that we carried out for the four post-periapse phases that deviate significantly from a blackbody spectrum imply that HD~80606b's atmosphere has an approximately solar metallicity with ${\rm [M/H]}=-0.17\pm0.31$---slightly below the host-star metallicity of $0.348\pm0.057$ \citep{rosenthal2021}---and a solar ${\rm C/O}=0.49\pm0.15$. We might expect that, as the planet approaches its host star, CH$_4$ is converted into CO through thermochemical reactions in response to the increase in temperature within the atmosphere (see Fig. \ref{fig:FpFs_PT} showing the derived PT profiles compared with the expected points along which CH$_4$ and CO are in equal abundance), however, in the case of HD~80606b, these reactions are expected to occur over timescales much longer than the orbital timescale \citep{iro2010,visscher2012,tsai2023}. The photochemical models generated for HD~80606b by \citet{tsai2023} predict that the abundances of CH$_4$, H$_2$O, and CO undergo significant changes in response to the increase in incident UV flux that occurs during periapse passage. The resulting change in CH$_4$ and CO abundance, along with that of several photochemical byproducts such as C$_2$H$_2$ and HCN, can differ by orders of magnitude from chemical equilibrium predictions that neglect photochemistry.

Using both detectors without including a flux offset, we find that CH$_4$ is detected in all four of the extracted post-periapse emission spectra while CO and H$_2$O are detected during two and three phases, respectively. We find that both the chemical equilibrium and free retrievals yield high-quality fits and that the latter models are statistically preferred due to the fewer number of free parameters. This suggests that the atmosphere is not strongly impacted by disequilibrium chemistry or photochemistry. Water is not detected during the first post-periapse phase, which results in a significantly super-solar ${\rm C/O}=0.74\pm0.015$ and ${\rm [M/H]}=1.24\pm0.65$ and a $2\sigma$ upper limit of the volume mixing ratio of $\log_{10} X_{\rm H_2O}<-3.1$ being inferred from the free retrieval. While this may be evidence of disequilibrium chemistry, it is ambiguous due to the fact that it is only associated with a single phase and may be related to the pre-/post-periapse transition from a blackbody spectrum. Ultimately, a more sophisticated modelling framework that can better capture shorter timescale variability and/or 3D effects may provide improved constraints on potential changes in the atmosphere's chemical composition.

\section{Conclusions}\label{sect:conclusions}
We present the first  partial phase curve observations of an exoplanet obtained with JWST. These NIRSpec/G395H measurements of HD~80606b's periapse passage reveal an atmosphere that undergoes significant temperature changes during the $21\,{\rm hr}$ observing window. This work demonstrate that partial phase curves with JWST can provide reliable data at twice the photon noise precision level. Time dependent spectral changes in the NIRSpec/G395H bandpass are observed in the emission spectrum of the planet during this observation. Our analysis suggests that, prior to periapse, the atmospheric layers probed at these wavelengths are predominantly isothermal based on the fact that no spectral features are detected and the derived planet spectrum is consistent with that of a blackbody. During the post-periapse phases, we detect CH$_4$, H$_2$O, and CO spectral absorption features with maximum significance levels of $10.7\sigma$, $5.5\sigma$, and $4.4\sigma$, respectively, when considering both the NRS1 and NRS2 detectors without including a flux offset parameter for NRS1. When including a flux offset, CH$_4$ and CO are the only species detected at $5.7\sigma$ and $3.8\sigma$, respectively, while considering only the NRS2 detector yields a $3.6\sigma$ detection of CO. We find that the post-periapse emission spectra are generally consistent with thermochemical equilibrium models and therefore, the atmosphere appears to be chemically mixed. No significant spectral emission features are detected at any of the observed phases, which, along with the PT profiles derived from atmospheric retrievals, rules out the presence of a predicted temperature inversion \citep{iro2010,lewis2017,mayorga2021,tsai2023}.

Considering the significantly higher precision of JWST NIRSpec observations relative to the \emph{Spitzer} $4.5\,{\rm \mu m}$ and $8\,{\rm \mu m}$ partial phase curves, additional measurements extending over a longer timespan may be capable of detecting a ringing effect caused by the planet's rotation/advection \citep{cowan2011,kataria2013,vanrespaille2024}. This would provide important and entirely unique constraints that are needed to accurately model the atmospheres of eccentric hot Jupiters like HD~80606b and hot Jupiters in general. Additional phase curve measurements of HD~80606b obtained at different wavelengths will also help to constrain the chemical and photochemical reactions that are likely occurring in the atmosphere. For example, by incorporating \emph{HST} WFC3/G141 measurements and/or \emph{JWST} NIRISS/SOSS or NIRSpec/G140M measurements, we gain sensitivity to gaseous H$_2$O \citep{constantinou2023,benneke2024}, which, similar to CH$_4$, is predicted to decrease in abundance with increasing CO due to photochemical reactions \citep{tsai2022}. Moreover, partial phase curves obtained at longer wavelengths using instruments like MIRI/LRS can provide insight into the composition and size of cloud particulates (e.g., the observed flux $\gtrsim20\,{\rm hrs}$ post-periapse is predicted to be significantly higher if MnS clouds are present compared to MgSiO$_3$ clouds) \citep{lewis2017}. High-resolution near IR observations have the potential to detect HCN \citep[e.g.,][]{gandhi2020e,sanchez-lopez2022a} that is expected to be found at high altitudes in hot Jupiter atmospheres \citep[e.g.,]{baeyens2024}. Therefore, future high-precision observations of HD~80606b and other eccentric hot Jupiters such as HAT-P-2b \citep{lewis2013}, HD~17156b \citep{kataria2013}, and XO-3b \citep{dang2022} using JWST or ground-based high-resolution spectrographs may provide important clues into the nature of hot Jupiter atmospheres and the impact of photochemistry and clouds.

\begin{table*}
	\caption{\textbf{Parameters derived for the NRS1 and NRS2 white light curves using the nominal models. For NRS1, we also list the parameters obtained when the $\Delta F_{\rm jump}$ parameter is not included in the model.} For each parameter, we report the median value calculated from the marginalized posteriors while the uncertainties correspond to 16$^{\rm th}$ and 84$^{\rm th}$ percentiles ($1\sigma$).}\vspace{-0.3cm}
	\label{tbl:NRS1_NRS2_WLC}
	\begin{center}
	\begin{tabular}{@{\extracolsep{\fill}}l c c r@{\extracolsep{\fill}}}
		\hline
		\hline
		\noalign{\vskip0.5mm}
Parameter & Nominal NRS1 & Nominal NRS1 & NRS2 \\
          &              & (No jump)    & \\
\hline
$F_0\,[{\rm e^-/s}]$ & $111550.08\pm0.80$ & $111554.08\pm0.60$ & $36531.35\pm0.23$\\ 
$c_1\,{\rm [ppm]}$ & $433.6\pm7.8$ & $466.3\pm8.0$ & $812.9\pm6.9$\\ 
$c_2-t_{\rm p}\,{\rm [hr]}$ & $0.638_{-0.080}^{+0.083}$ & $0.613_{-0.072}^{+0.068}$ & $1.073_{-0.039}^{+0.040}$\\ 
$c_3\,{\rm [hr]}$ & $5.42\pm0.18$ & $4.62\pm0.11$ & $7.65\pm0.10$\\ 
$T_{\rm e}[{\rm BJD}-2459885]$ & $0.16528\pm0.00025$ & $0.16487\pm0.00027$ & $0.16421\pm0.00016$\\ 
 
\hline

$A_1\,{\rm [ppm]}$ & $8.1_{-4.2}^{+4.5}$ & $8.5_{-5.1}^{+4.7}$ & $2.0_{-1.4}^{+2.8}$\\ 
$B_1\,{\rm [ppm/hr]}$ & $2.50\pm0.36$ & $2.79_{-0.39}^{+0.41}$ & $2.87\pm0.29$\\ 
$\phi_1$ & $0.083\pm0.011$ & $0.0427\pm0.0091$ & $0.831\pm0.012$\\ 
$f_1\,{\rm [hr^{-1}]}^\dagger$ & 0.174 & 0.174 & 0.231\\ 

\hline

$p_1\,{\rm [ppm/hr]}$ & $-153.74\pm0.76$ & $-148.92_{-0.44}^{+0.48}$ & \\ 
$p_2\,{\rm [ppm/hr^2]}$ & $3.395_{-0.080}^{+0.078}$ & $3.651_{-0.078}^{+0.063}$ & \\ 
$c_y\,[\rm ppm]$ & $-2.6\pm1.4$ & $-2.3\pm1.7$ & $-0.6\pm1.9$\\ 
$c_{{\rm GS,}y}\,[\rm ppm]$ & $2.6\pm1.4$ & $3.0\pm1.6$ & $11.5\pm1.9$\\ 
$T_{\rm jump}\,[{\rm BJD}-2459885]$ & $0.16675_{-0.00058}^{+0.00042}$ &  & \\ 
$\Delta(F_{\rm p}/F_\star)_{\rm jump}\,[\rm ppm]$ & $77\pm10$ &  & \\ 
$\log({\rm Jit}/{\rm e^-/s})$ & $2.7821_{-0.0080}^{+0.0079}$ & $2.7859_{-0.0078}^{+0.0093}$ & $1.9020_{-0.0082}^{+0.0086}$\\ 
\hline
		\noalign{\vskip0.5mm}
		\hline
\multicolumn{4}{l}{$^\dagger$Fixed at values inferred from residuals calculated for the simplest considered} \\
\multicolumn{4}{l}{light curve model (see Fig. \ref{fig:centroid}).}\\
	\end{tabular}
	\end{center}
\end{table*}

\acknowledgments{JTS and JL acknowledge support from NSF AAG award AST2009343. JTS acknowledges funding support from grant 22JWGO1-17 awarded by the CSA. This work is based on observations made with the NASA/ESA/CSA James Webb Space Telescope. The data were obtained from the Mikulski Archive for Space Telescopes at the Space Telescope Science Institute, which is operated by the Association of Universities for Research in Astronomy, Inc., under NASA contract NAS5-03127 for JWST. These observations are associated with program \#2488. KDC acknowledges support for program \#2488 was provided through a grant from the STScI under NASA contract NAS5-03127. J.M.D acknowledges support from the Amsterdam Academic Alliance (AAA) Program, and the European Research Council (ERC) European Union’s Horizon 2020 research and innovation program (grant agreement no. 679633; Exo-Atmos). This work is part of the research program VIDI New Frontiers in Exoplanetary Climatology with project number 614.001.601, which is (partly) financed by the Dutch Research Council (NWO).  L.D.\ acknowledges support from the Banting Postdoctoral Fellowship program, administered by the Government of Canada and the Trottier Family Foundation.

We thank Everett Schlawin, Shang-Min Tsai, and Maria Steinrueck for helpful discussions that improved the analysis of the data and the interpretation of the results presented in this work.}

All the {\it JWST} data used in this paper can be found in MAST: \dataset[10.17909/yb30-hj11]{http://dx.doi.org/10.17909/yb30-hj11}.

\facilities{JWST (NIRSpec)}

\software{\texttt{AstroPy} \citep{astropycollaboration2013,astropycollaboration2018,astropycollaboration2022}, \texttt{matplotlib} \citep{Hunter:2007}, \texttt{numpy} \citep{harris2020array}, \texttt{scipy} \citep{2020SciPy-NMeth}, \texttt{petitRADTRANS} \citep{molliere2019}, \texttt{Eureka!} \citep{bell2022}, \texttt{jwst}, \texttt{batman} \citep{kreidberg2015}, \texttt{emcee} \citep{foreman-mackey2013,foreman-mackey2019}, easyCHEM \citep{molliere2017}, \texttt{PyratBay} \citep{cubillos2021}.}

\clearpage

\section*{Appendix}\label{sect:appendix1}

\begin{figure}
	\centering
    \includegraphics[width=0.8\columnwidth]{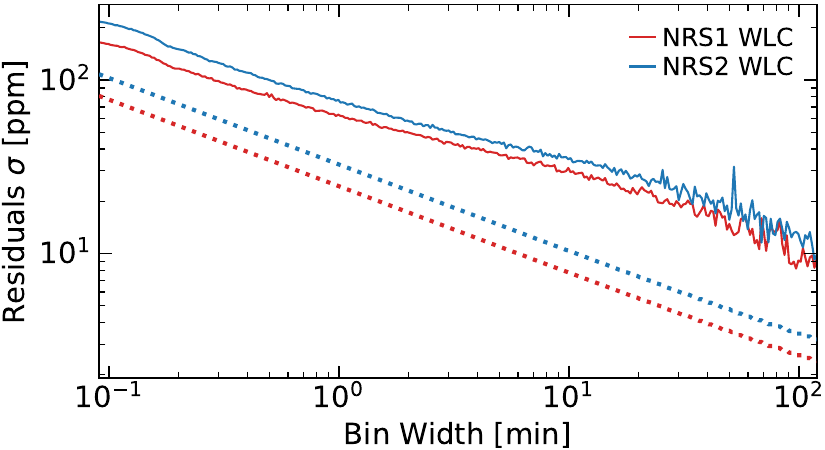}
    \includegraphics[width=0.8\columnwidth]{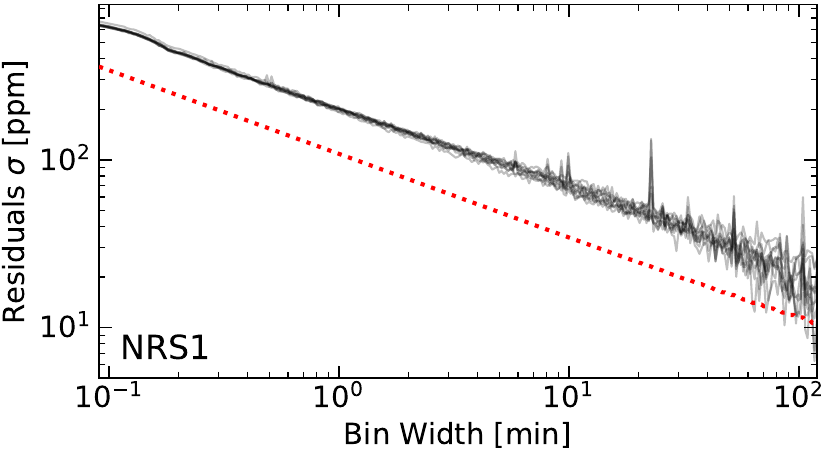}
    \includegraphics[width=0.8\columnwidth]{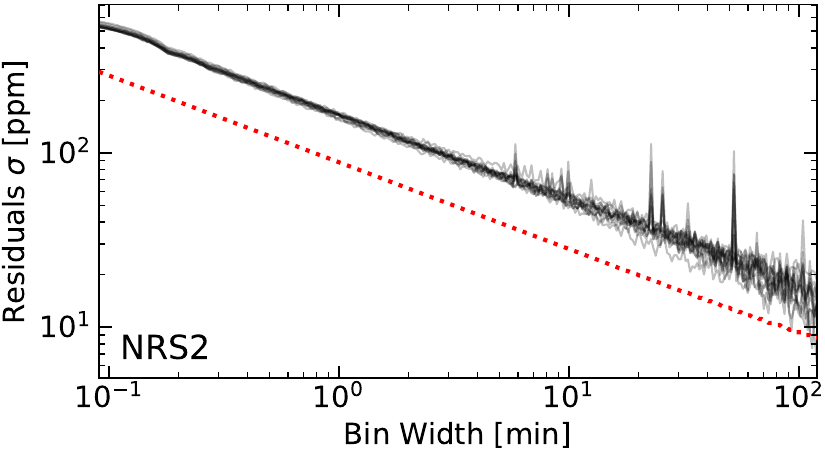}
	\caption{\emph{Top:} Allan deviation plots for the NRS1 (solid blue) and NRS2 (solid red) white light curves. Dashed lines correspond to the expectations from pure photon noise ($81\,{\rm ppm}$ for NRS1 and $108\,{\rm ppm}$ for NRS2); the unbinned light curves have a scatter of $2.03\times$ and $2.02\times$ the photon noise. \emph{Middle and bottom:} Allan deviation plots for the 10 NRS1 spectral light curves and the 14 NRS2 spectral light curves normalized to the values associated with the first wavelength channel in order to compare across all channels. The spectral light curves have a scatter of $1.71-1.94\times$ each channel's photon noise.}
	\label{fig:NRS1_NRS2_SLC_allan}
\end{figure}

\begin{figure*}
	\centering
     \includegraphics[width=1\textwidth]{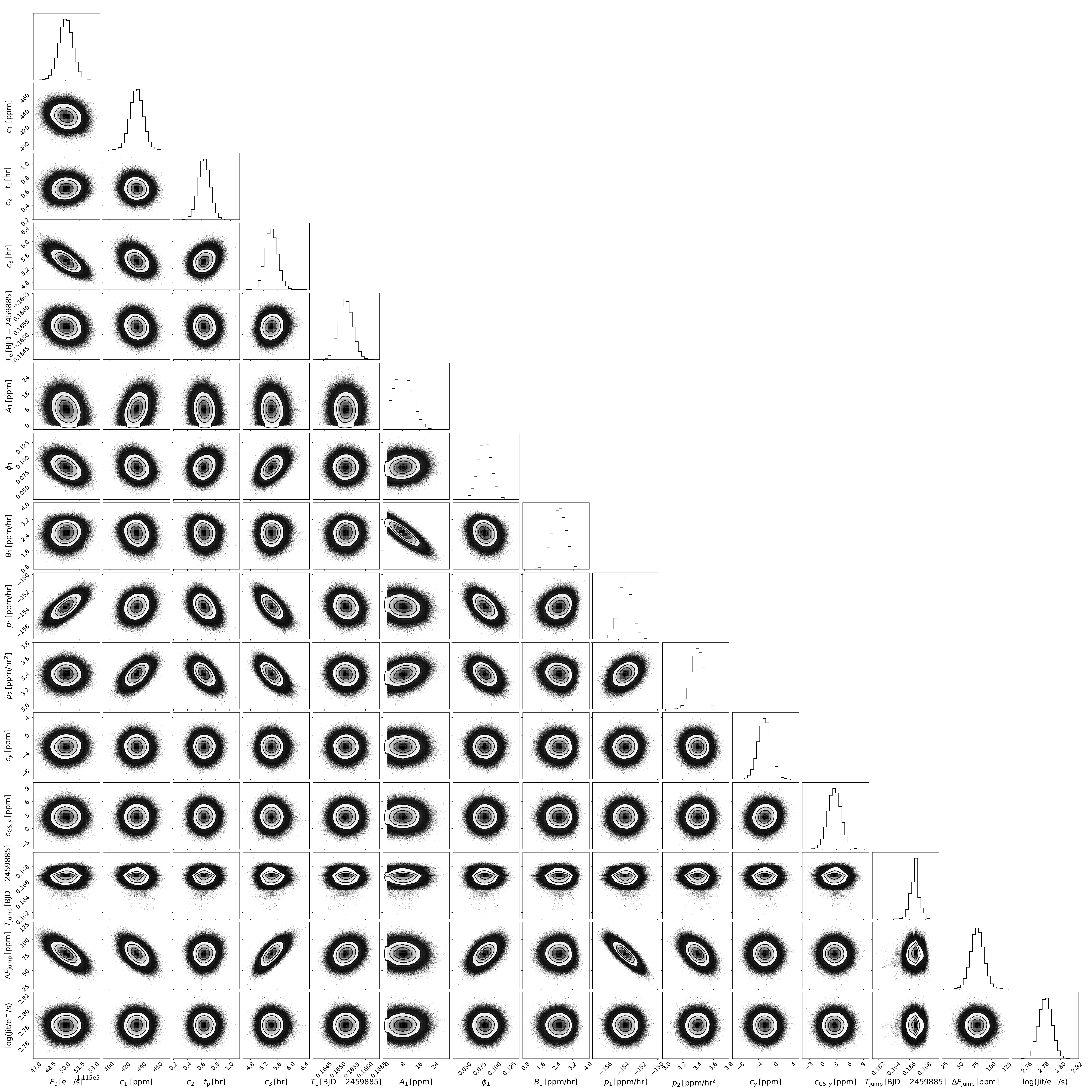}
	\caption{Marginalized posterior distributions for the NRS1 white light curve using the nominal model.}
	\label{fig:nrs1_wlc_corner}
\end{figure*}

\citet{moran2023} show that the two NIRSpec/G395H transit observations of GJ486b (GO-1981, PI:Stevenson) exhibit strong, wavelength dependent slopes similar to what we find in our analysis of the NRS1 measurements of HD~80606b. This is evident when comparing the 2D spectra shown in our Fig. \ref{fig:NRS1_NRS2_2D} with Fig. 1 of \citet{moran2023}. Here we present a more quantitative comparison between the linear and quadratic trends associated with both NRS1 data sets.

\begin{figure}
	\centering
    \includegraphics[width=1\columnwidth]{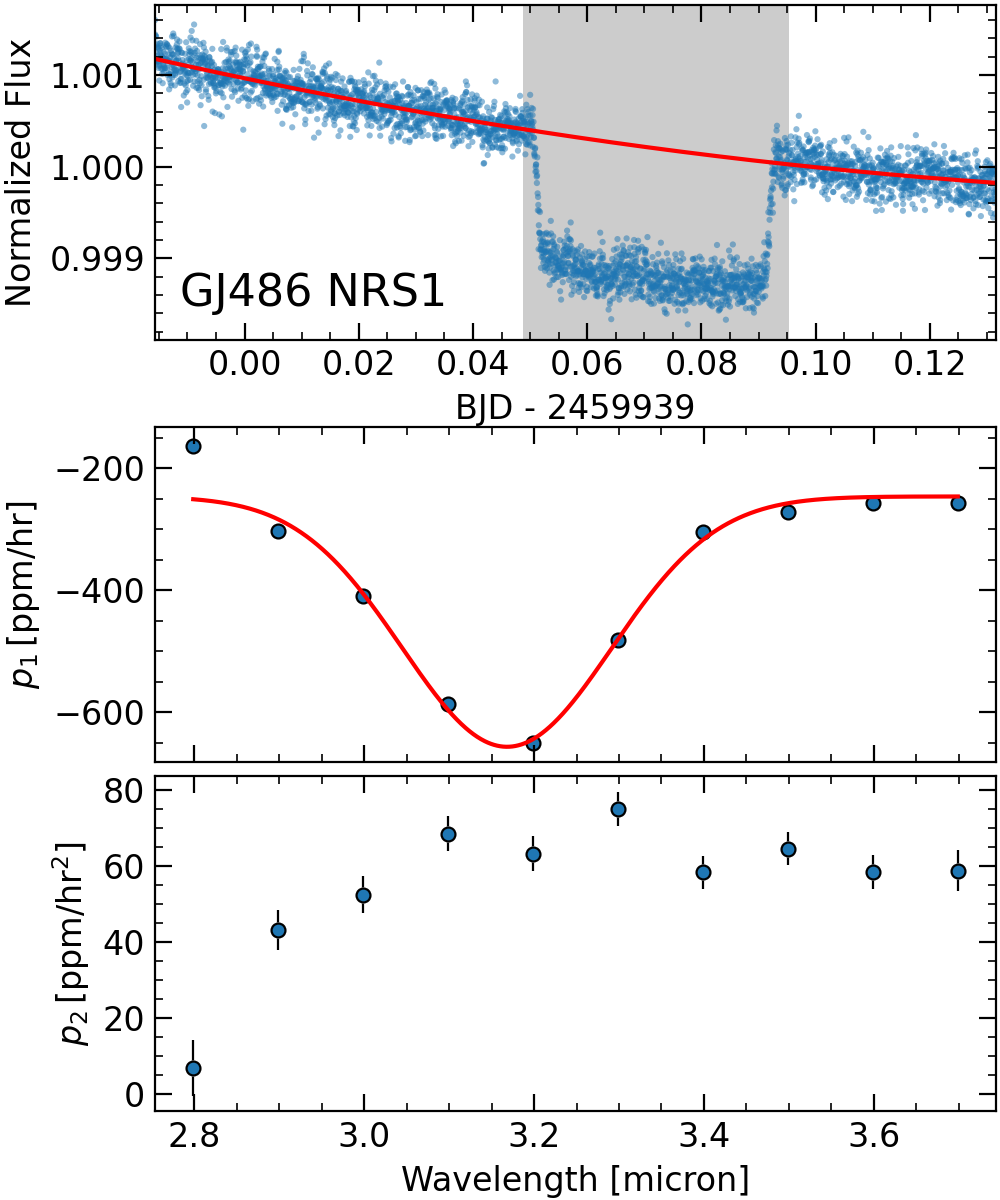}
	\caption{Out-of-transit baseline fit to the first of two publicly available NRS1 transit observations of GJ486b, which was obtained using NIRSpec/G395H with 3 groups. The top panel shows the white light curve with the associated $2^{\rm nd}$ order polynomial baseline  model (red); the out-of-transit measurements (blue points outside the gray box) have residuals with a $153\,{\rm ppm}$ scatter. The middle and bottom panels show the linear ($p_1$) and quadratic ($p_2$) baseline polynomial terms derived from the spectral light curves. The red curve in the middle panel is a Gaussian fit used for comparison with the similar trend found in the HD~80606 NRS1 observation.}
	\label{fig:gj486b_p1p2}
\end{figure}

\begin{figure*}
	\centering
     \includegraphics[width=1\textwidth]{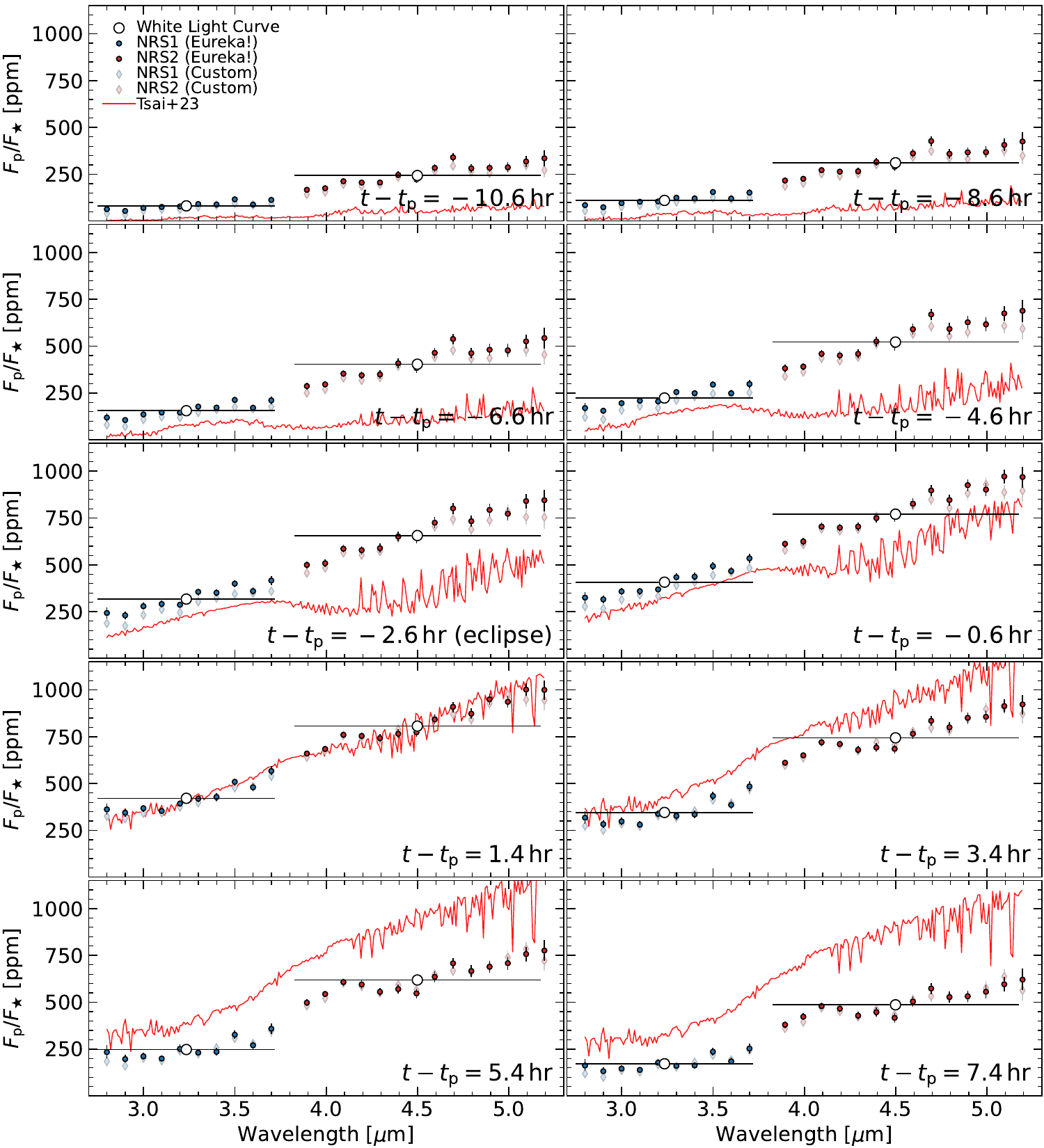}
	\caption{Planet-to-star flux contrast derived using the NRS1 and NRS2 light curve models for the \texttt{Eureka!} reduction (dark circles) and the custom reduction (light diamonds). Red lines show the model spectra published by \citet{tsai2023} after applying instrumental broadening and binning for visual clarity.}
	\label{fig:FpFs_spec}
\end{figure*}

We reduced the NRS1 G395H time series observations for the first of two transit observations of GJ486b using essentially the same \texttt{Eureka!} control files used for the HD~80606b data. We used an aperture half-width of 5 pixels rather than 4 (as adopted for the HD~80606 reduction) to match what was used by \citet{moran2023}. We also manually masked 9 columns (i.e., wavelength channel light curves) that were identified by eye as having anomalous flux values. We then extracted the white light curve and 10 spectral light curves and fit a $2^{\rm nd}$ order polynomial to each light curve's out-of-transit baseline measurements. The baseline fit to the white light curve is shown in Fig. \ref{fig:gj486b_p1p2}; the residuals for this baseline fit have a $153\,{\rm ppm}$ scatter, which is comparable to the $132-158\,{\rm ppm}$ scatter obtained by \citet{moran2023} for the same data set using \texttt{Eureka!}, \texttt{FIREFLy} \citep{rustamkulov2022}, and \texttt{Tiberius} \citep{kirk2018,kirk2021} reduction pipelines.

The polynomial coefficients, $p_1$ and $p_2$, derived from the GJ486 NRS1 spectral light curves are shown in the bottom two panels of Fig. \ref{fig:gj486b_p1p2}. We find that the $p_1(\lambda)$ values show a similar Gaussian-like trend as that obtained for HD~80606's NRS1 data set with minimum values occurring at the $3.2\,{\rm \mu m}$ wavelength bin. The GJ486 results show a $p_2(\lambda)$ that is essentially constant for $\lambda\geq3.1\,{\rm \mu m}$ and which decreases for the 3 $\lambda<3.1\,{\rm \mu m}$ bins. We note that the GJ486 baseline is likely impacted by stellar activity/star spots, which, along with the different instrument setups, prevent a one-to-one comparison with the HD~80606 polynomial coefficients from being carried out.

\begin{figure*}
	\centering
     \includegraphics[width=1\textwidth]{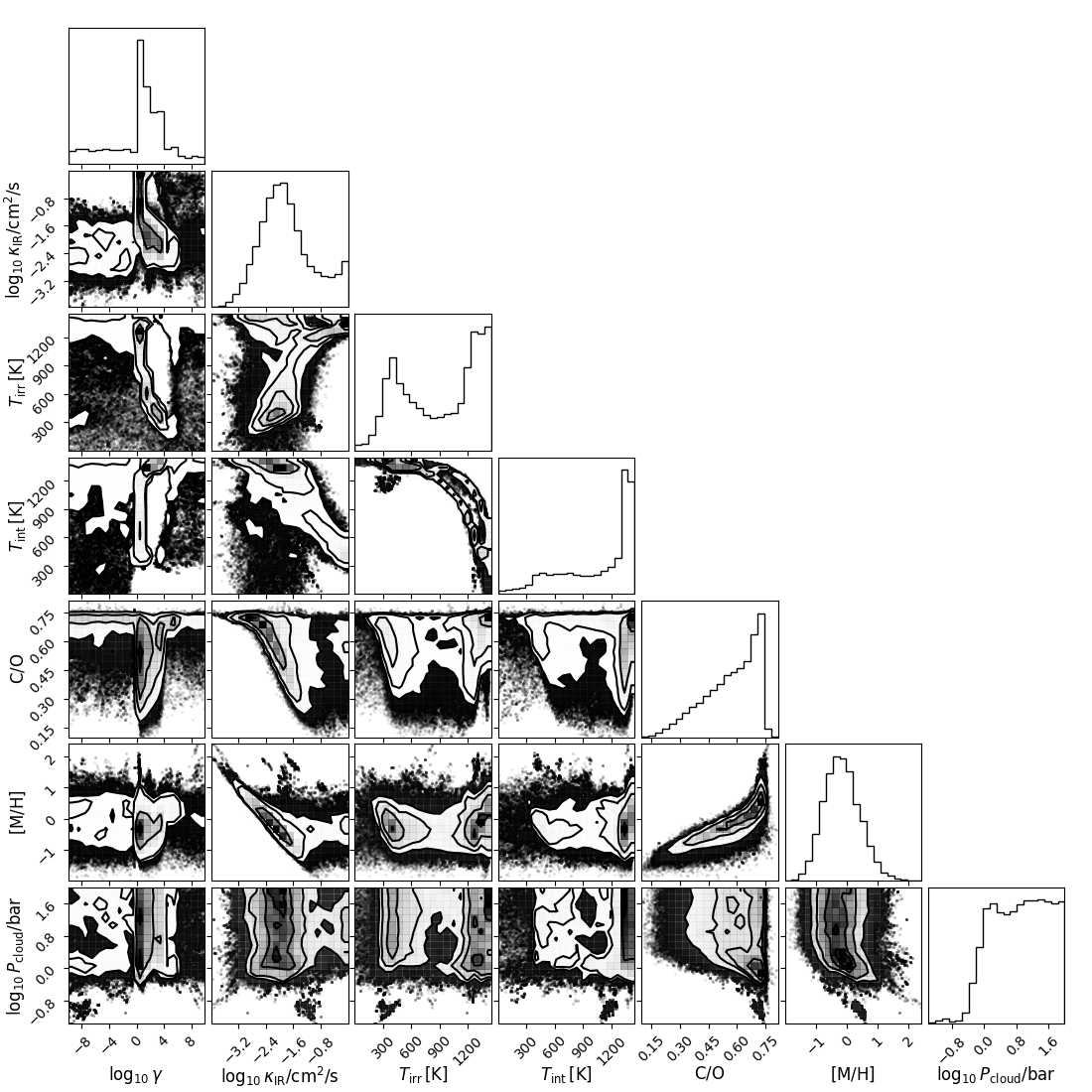}
	\caption{Marginialized posteriors derived for the chemical equilibrium retrieval for the $t-t_{\rm p}=3.4\,{\rm hr}$ phase shown in Fig. \ref{fig:FpFs_PT}.}
	\label{fig:chem_eq_corner}
\end{figure*}

\clearpage

\onecolumngrid

\begin{figure*}
    \centering
     \includegraphics[height=5.4cm]{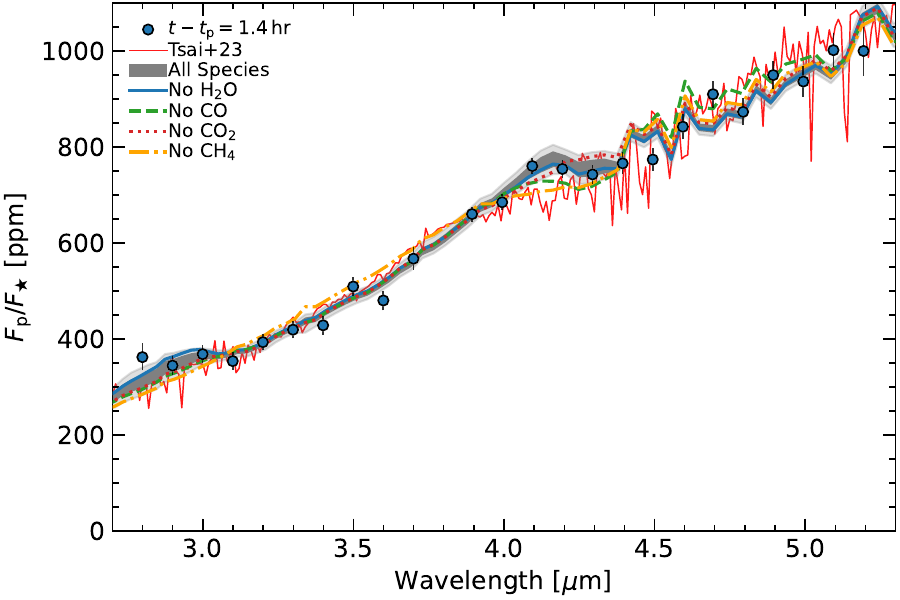}
     \includegraphics[height=5.8cm]{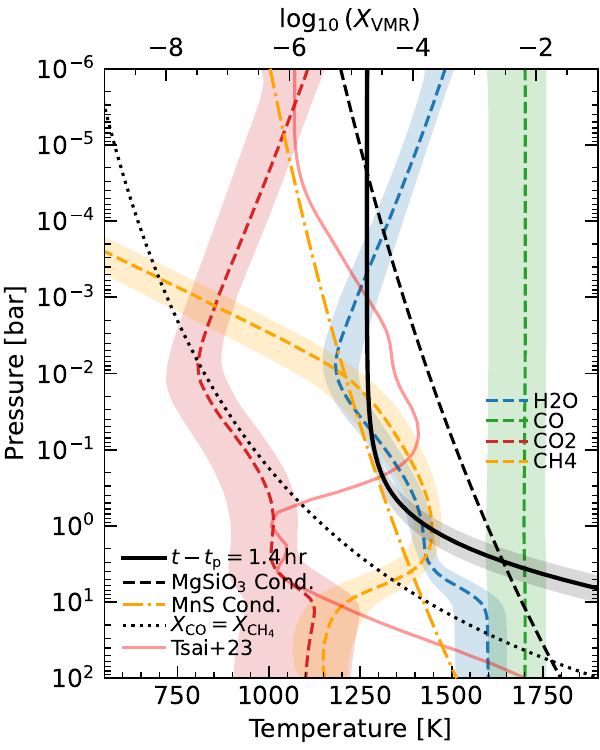}

     \includegraphics[height=5.4cm]{figures/spectrum/HD80606b_single_7_ChemEq_G2010_Pcloud_7_FpFs_spec.pdf}
     \includegraphics[height=5.8cm]{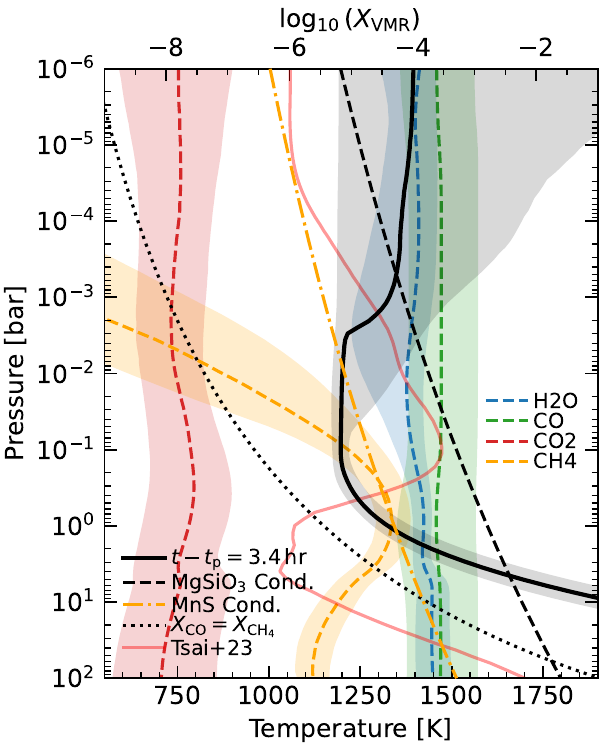}

     \includegraphics[height=5.4cm]{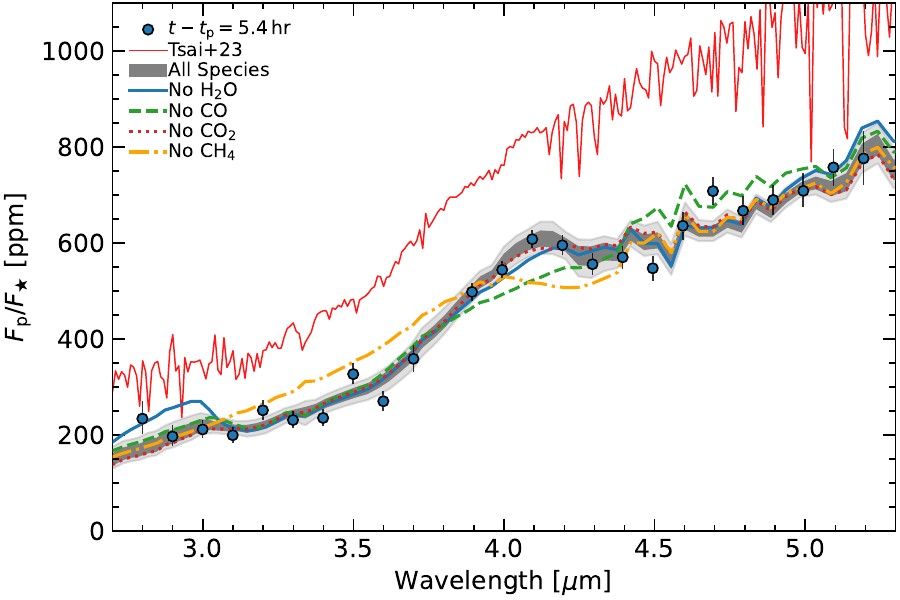}
     \includegraphics[height=5.8cm]{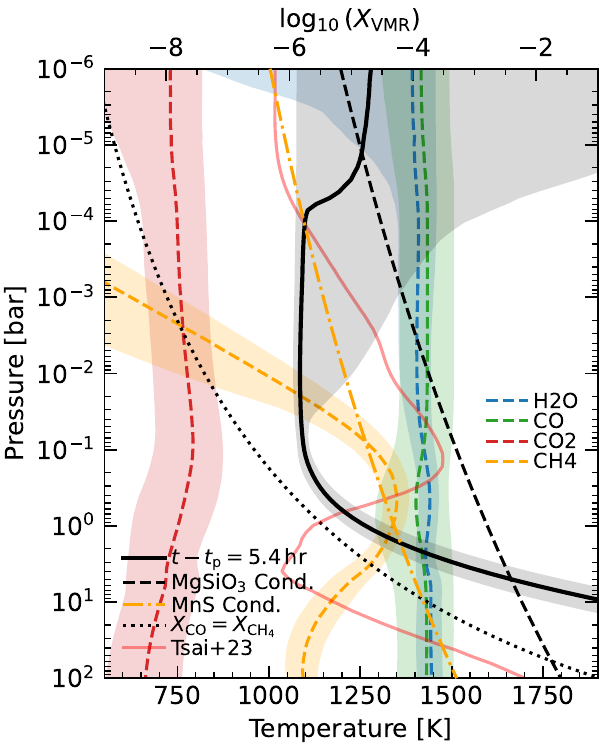}

     \includegraphics[height=5.4cm]{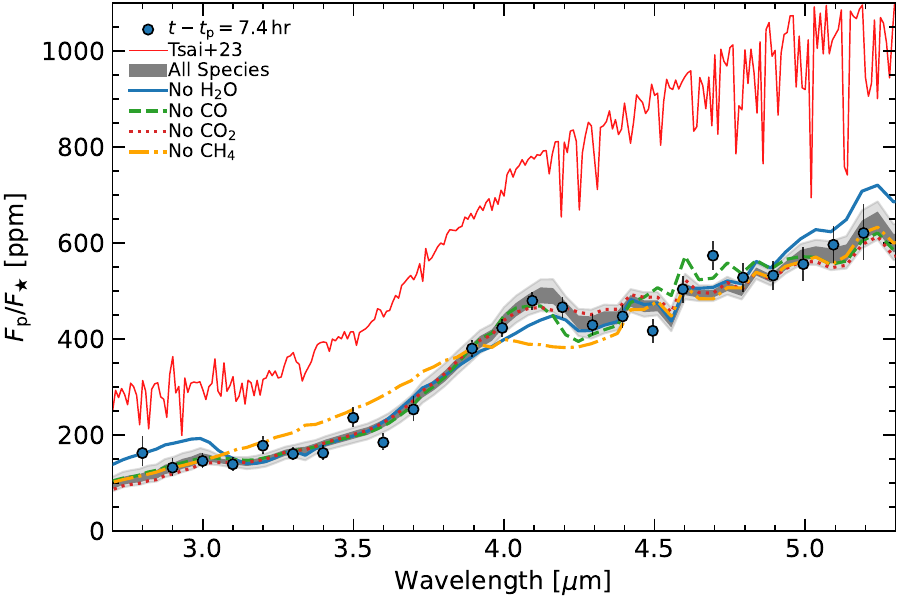}
     \includegraphics[height=5.8cm]{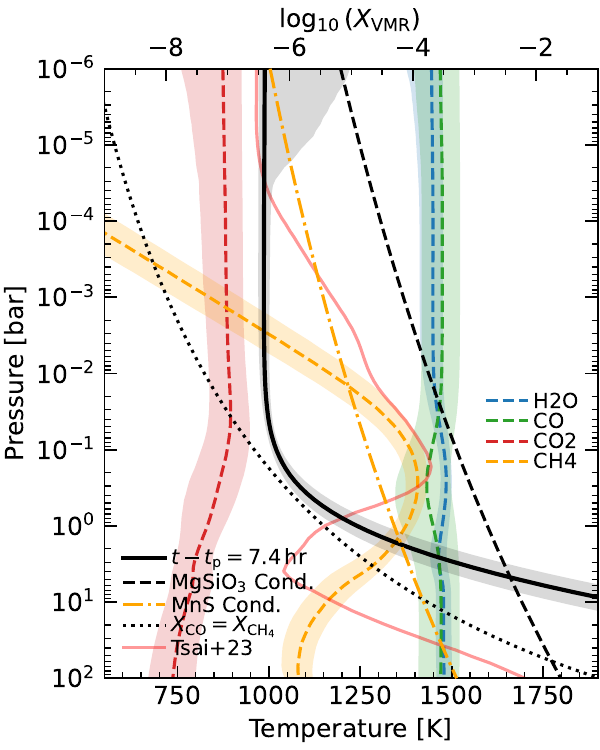}
    \caption{Chemical equilibrium retrieval results (similar to Fig. \ref{fig:FpFs_PT}) for the last four phases where the PT profiles include the H$_2$O, CO, CO$_2$, and CH$_4$ abundances plotted as VMRs along the top axis. All colored contours correspond to $1\sigma$.}
    \label{fig:FpFs_PT_all}
\end{figure*}

\twocolumngrid

\bibliography{HD80606b}{}
\bibliographystyle{aasjournal}

\end{document}